\documentclass[12pt]{article}

\usepackage{hyperref}

\usepackage{epsf}
\usepackage{amsmath,amssymb}
\usepackage{latexsym}
\usepackage{graphicx,color}
\usepackage{wrapfig}
\usepackage{latexsym}
\usepackage{graphics}

\usepackage{color}
\usepackage{delarray,color}
\usepackage{fancybox}  

\newcommand {\beq}{\begin{align}}
\newcommand {\eeq}{\end{align}}

\newcommand{\de}{\partial}
\newcommand{\be}{\begin{equation}}
\newcommand{\ba}{\begin{align}}
\newcommand{\ea}{\end{align}}
\newcommand{\ee}{\end{equation}}

\newcommand{\z}{{\mathbf Z }}
\newcommand{\R}{{\mathbf R }}

\newcommand{\beqa}{\begin{align}}
\newcommand{\eeqa}{\end{align}}
\newcommand{\CR}{\nonumber \\}
\newcommand{\pa}{\partial}

\newcommand{\cO}{{\cal O}}

\newcommand{\unit}{\hbox to 3.8pt{\hskip1.3pt \vrule height 7.4pt
    width .4pt \hskip.7pt \vrule height 7.85pt width .4pt \kern-2.4pt
    \hrulefill \kern-3pt \raise 3.7pt\hbox{\char'40}}}

\def\matt[#1,#2,#3,#4]{\left(%
\begin{array}{cc} #1 & #2 \\ #3 & #4 \end{array} \right)}

\newcommand{\ket}[1]{{\left| #1 \right\rangle}}
\newcommand{\bra}[1]{{\left\langle#1\right|}}
\usepackage{tabularx}

\catcode`\@=11
\@addtoreset{equation}{section}

\catcode`@=12
\relax

\begin{document}

\begin{titlepage}

\setcounter{page}{0}

\renewcommand{\thefootnote}{\fnsymbol{footnote}}

\begin{flushright}
YITP-19-62
\end{flushright}

\vskip 1.35cm

\begin{center}
{\Large \bf 
Classical Limit of 
Large $N$ Gauge Theories with Conformal Symmetry
}

\vskip 1.2cm 

{\normalsize
Seiji Terashima\footnote{terasima(at)yukawa.kyoto-u.ac.jp}
}

\vskip 0.8cm

{ \it
Yukawa Institute for Theoretical Physics, Kyoto University, Kyoto 606-8502, Japan
}

\end{center}

\vspace{12mm}

\centerline{{\bf Abstract}}

In this paper 
we study classical limit of conformal field theories
realized by large $N$ gauge theories 
using the generalized coherent states.
For generic large $N$ gauge theories with conformal symmetry, 
we show that 
the classical limit of them 
is described by the classical Einstein gravity.
This may be regarded as a kind of derivation of the AdS/CFT correspondence.

\end{titlepage}
\newpage

\tableofcontents
\vskip 1.2cm 

\section{Introduction and summary}

According to the $AdS/CFT$ correspondence \cite{Maldacena},
a certain class of 
$d$-dimensional conformal field theories ($CFT_d$)  
correspond to 
$d+1$-dimensional quantum gravity theory on 
an asymptotically $AdS_{d+1}$ spacetime.
This conjecture has been investigated intensively
and there are many evidences for this conjecture,
although there is no proof. 
The explicit relation of the $AdS/CFT$ correspondence
is the GKPW relation \cite{GKP, W}
where 
a $CFT$ partition function with the source terms
is identified with 
the partition function of a quantum gravity 
on $AdS$ with appropriate boundary conditions corresponding 
to the source terms.
The extrapolation formula \cite{BDHM},
which state that the boundary value of the bulk field is the $CFT$ primary field,
can be used as an explicit relation between the two theories.
These are the ``dictionaries'' of the AdS/CFT correspondence and the basic assumptions
of the most of the studies.

Alternatively, we can say that 
the $AdS/CFT$ correspondence
is the equivalence of the two theories as a quantum theory.
More explicitly,
the correspondence means that 
the Hilbert spaces and the Hamiltonians of 
the two theories 
in the operator formalism
are equivalent.\footnote{
For this, we need to choose a time direction
and the usual choice is the $CFT_d$ 
on ${\mathbf R} \times S^{d-1}$ where 
${\mathbf R} $ represents the time.
Note that in this formalism we can consider the sourceless case, which is simpler.}
In this formulation, 
a proof of the AdS/CFT correspondence means 
showing the spectrum of $CFT_d$ 
is equivalent to the spectrum of a quantum gravity
on asymptotic $AdS_{d+1}$.

In \cite{op},
instead of assuming such dictionaries or existence of 
a bulk dual, 
we studied the (low energy) spectrum of
generic large $N$ gauge theories with conformal symmetry, in the leading order in the large $N$ limit,
and found that 
it is identical 
to the spectrum of the Einstein gravity theory in global $AdS_{d+1}$ in the free theory limit, 
following some earlier works 
\cite{BKL, BDHM, Pol} (see also \cite{FK}-\cite{Rehren}).
Here, ``generic'' means that
the theories satisfy the following two properties.
The first one is that 
the low energy spectrum is 
determined only by the conserved symmetry currents
whose conformal dimension is protected against the quantum corrections.\footnote{
This is similar to the hydrodynamics.
}
In this paper, we further assume that the symmetry of the $CFT_d$ is the conformal symmetry only, for simplicity.
The second one is that the states generated from the 
symmetry currents acting on the vacuum
are completely independent except the relations imposed by the symmetry.\footnote{
Conversely, this complete independence is needed 
for the $CFT_d$
to have a gravity dual, in the limit.
}
These properties are highly expected
for the large $N$ strongly coupled gauge theories.\footnote{
Here, the theories with weak t'Hooft coupling is not
regarded as a generic theories.
}
With this explicit identification of the spectrum of $CFT$
to the spectrum of the gravity on $AdS$ space,
we derived 
the GKPW relation.

In this paper, we include the $1/N$ corrections
to the study of \cite{op}, which are expected to correspond 
to the interactions in the gravity side, although 
we do not assume the existence of the gravity dual 
as in \cite{op}.
What we assume is the above two properties 
and the large $N$ factorization which is certainly satisfied 
for the t'Hooft
large $N$ limit of gauge theories in the leading order in $N$.\footnote{
More explicitly, we will assume that only 
the energy momentum tensor is the low energy 
primary field which has large $C_T$ in (\ref{opet})
and the states generated by this
are completely independent for the energy below some $C_T$-dependent
large energy scale.
This $C_T$ plays the role of $N^2$.
}
Instead of including $1/N$ corrections order by order,
we will consider a large $N$ limit different form the large $N$ limit
taken in \cite{op} corresponding to free theory,
which we will call the naive large $N$ limit below.
This large $N$ limit considered in this paper corresponds 
to the the classical limit of the theory, although there is 
no parameter like $\hbar$ in general a CFT, and
the role of $\hbar$ is played by $1/N^2$.
In the gravity dual, 
the classical limit we consider in this paper, 
$1/N^2  \rightarrow 0$ is indeed realized by 
$G_N \rightarrow 0 $
because $1/G_N$  is an overall factor of the action
and $1/G_N$ will be identified as $N^2$.

More explicitly, we would like to understand
what is
the classical limit of the generic large $N$ gauge theories 
with conformal symmetry.
In order to answer this,
an important object is the algebra of the 
energy momentum tensor.
This is an analogue of the Virasoro algebra 
for $d>2$.
As conserved charges of the theory,
an analogue of the Virasoro algebra is 
the algebra of the conformal symmetry generators.
However, if we regard the generators of the Virasoro algebra
as modes of the energy momentum tensor,
we can consider the commutator algebra of the modes of the 
energy momentum tensor for $d>2$.
Because of the assumptions, the generators of 
this algebra spans all operators on the low energy theories.
Thus, the classical limit of the theory is given by
the classical limit of this algebra, which gives the Poisson bracket.
It is important to note that this algebra is almost unique,
like Virasoro algebra, and the Brown-York tensor (or the boundary stress tensor) of classical gravity on asymptotic $AdS_{d+1}$ also forms
the algebra by the Poisson bracket.
We will show that 
the classical limit of the generic large $N$ gauge theory with conformal symmetry 
is 
the classical dynamics of the Einstein gravity on asymptotic $AdS_{d+1}$
by showing that the Poisson bracket and the Hamiltonian of the CFT 
are equivalent to the those of the gravity.
This may be regarded as a kind of derivation of the AdS/CFT correspondence from CFT.

Conversely, we can say that 
the classical dynamics of gravity on asymptotic $AdS_{d+1}$ is 
a classical dynamics of the energy momentum tensor
of $CFT_d$.
Thus, 
the quantization of gravity can be considered as 
finding a quantum system which has an
``energy momentm tensor'' which becomes
the energy momentum tensor of CFT for a large central charge $C_T$
 in the classical limit.
(We also require that it has the generic spectrum
in the large $C_T$ limit.)
Of course, the natural choice for such theory 
is the CFT and the tensor as the energy momentum tensor itself.
In general, the (classical) dynamics of an appropriate system in low energy limit will be
described by the hydrodynamics in which only the energy momentum tensor appears.
Because only the energy momentum tensor of the CFT appears 
in the classical gravity,
this might mean that the classical gravity is a kind of thermal physics.

There are many things we do not understand in this paper.
In particular, the black holes in the brick wall picture
\cite{tHooft, brick} in the classical dynamics obtained in this paper will be interesting to be investigated. We hope to report this in near future.

This paper is organized as follows.
In the next section,
we study the algebra of the energy momentum tensor by 
expanding it on the cylinder.
In section three, 
we consider the classical limit of the $CFT_d$
using generalized coherent states.
It is shown that the classical limit of the $CFT_d$
is 
the classical dynamics of the Einstein gravity on asymptotic $AdS_{d+1}$
in the final section. 
In the appendix, relation between the classical limit of AdS/CFT and the naive large $N$ limit is explained
and some discussions on generalized coherent states are given.

\section{Algebra for the energy momentum tensor}

In this section,
we will expand 
the energy momentum tensor 
$T_{\mu \nu}$ of the $CFT_d$  on the cylinder ${\mathbf R} \times S^{d-1}$
by the ``spherical harmonics'' of $S^{d-1}$
and the energy, which is the eigen value of the dilatation $D$,
to the infinitely many operators.
These operators are analogues of the generators of the 
Virasoro algebra of the $CFT_2$.
We will divide these operators to three classes
(positive, negative and isotropy), which we will explain.
Then, the commutator algebra of them will be studied from 
the operator product expansion (OPE).
Because 
the low energy states are spanned by these operators
acting on the vacuum,
the algebra is considered as the operator algebra
of the low energy theory of the generic large $N$ 
gauge theory with conformal symmetry.

For a review of the $CFT_d$, see, for example, \cite{SD}.

\subsection{Scalar case}

First, as a warm-up example, 
let us consider the (normalized) scalar primary field $\cO_\Delta(x)$, instead of $T_{\mu \nu}$, on $R^d$.
Here,
we regard the operators are defined by the radial quantization,
thus, it is expanded by the spherical harmonics and the radial direction $|x|=\sqrt{\sum_{\mu=1}^d x^\mu x^\mu}$ as
\begin{align}
\cO_\Delta (x)=\sum_{l=0}^\infty \sum_{m=0}^{m_{max}(l)}
Y_{l m} (\Omega) 
\cO_{\Delta  lm } (|x|),
\label{se1}
\end{align}
where $\cO_{\Delta  lm } (|x|) = \sum_\omega |x|^{\omega-\Delta} \cO_{\Delta \omega lm }$ and 
$\cO_{\Delta \omega lm}$ is the operator acting on the state of the theory on the $S^{d-1}$,
whose coordinates are denoted by $\Omega$, 
and $\omega$ is the energy in the cylinder coordinates
which is the eigen value of the dilatation $D$.The energy  $\omega$ will take continuous values.
Note that 
in the naive large $N$ limit taken in \cite{op},
\begin{align}
n=(|\omega|  -\Delta-l)/2,
\label{ln}
\end{align}
should be half integer because
operators which violate this integer condition gives states which can not be 
in the states spanned by 
the descendant states of the primary operator $\cO_\Delta(x)$.
Including the $1/N$ corrections,
this $n$ will be slightly modified by the corrections.
Moreover, the theory is not free theory except the naive large $N$ limit,
$\omega$ can take any value which is the difference between 
the energies of the states of the theory.
Here the spectrum of the states are almost the Fock space of the free theory,
but it is modified by the $1/N$ corrections.
\footnote{
For the Virasoro algebra, $|\omega| =l$, which is an integer, because of 
the traceless condition and 
the current conservation, which is imposed by the e.o.m. There are no constraints 
like this for the higher dimensional case.
}
(Here, the conformal weight $\Delta$ is the one includes the $1/N$ corrections.)

In (\ref{se1}), $m_{max}(l)$ is the number of the independent spherical harmonics, which depends on $d$ and $l$.  
Here, this scalar field is assume to be Hermite. 
The scalar field on the cylinder, 
$ds^2=d \tau^2+d \Omega^2$ where $\tau=\log r$,
is given by 
\begin{align}
\cO_\Delta^{cy} (\tau, \Omega)=\sum_{l=0}^\infty \sum_{m=0}^{m_{max}(l)}
Y_{l m} (\Omega) \sum_\omega 
e^{\omega \tau} \cO_{\Delta \omega lm },
\end{align}
then, the reflection positivity on the cylinder (or the usual Hermite conjugate on the Lorentzian cylinder) requires
\begin{align}
\cO_{\Delta \omega lm}^\dagger =  \cO_{\Delta (-\omega )l m},
\end{align}
where we take $Y_{lm}$ real.

Because $\cO_\Delta(x) \ket{0}$ should be regular for $x=0$
and $Y_{l m} (\Omega) |x|^{l}$ is a polynomial of $x^\mu$,
we see that $\cO_{\Delta \omega  lm} \ket{0}=0$ for $\omega <  \Delta+l$.\footnote{
We also see that if $n$ defined in (\ref{ln}) is not a half-integer, 
$\cO_{\Delta \omega  lm} \ket{0}=0$
by imposing the regularity of $(\partial_x^2)^m  \cO_\Delta(x) \ket{0}$
with a sufficiently large integer $m$.
}
Similarly, the Hermite condition implies that  $\bra{0}  \cO_{\Delta \omega lm} =0$ for $\omega  > -( \Delta+l)$.
Thus the operators $\cO_{\Delta \omega lm}$ 
with $|\omega | < \Delta+l$ satisfy 
$0=\cO_{\Delta \omega lm}\ket{0} =\bra{0}  \cO_{\Delta \omega lm}$
and we will denote these operators 
as $\cO^{iso}_{\Delta}$.
Note that the the commutator of
arbitrary two operators in $\cO^{iso}_{\Delta}$
also satisfies 
$[\cO_{\Delta \omega  lm}, \cO_{\Delta \omega ' l' m' }] \ket{0}
=\bra{0}  [\cO_{\Delta \omega  lm}, \cO_{\Delta \omega ' l' m' }]=0$.
This $\cO^{iso}_{\Delta}$ plays no roles 
in the naive large $N$ limit taken in \cite{op}
except the generators of the conformal group
which are in $\cO^{iso}_{\Delta}$ for the energy momentum tensor.

We will denote the operators $\cO_{\Delta \omega  lm}$ 
with $\omega   \geq \Delta+l$ and $\omega   \leq -(\Delta+l)$ 
as $\cO^{+}_{\Delta}$  and  $\cO^{-}_{\Delta}$, respectively.
The operators in $\cO^{+}_{\Delta}$ correspond to
the creation operators of the free scalar theory on 
the AdS background in the naive large $N$ limit \cite{op}, however, 
the notation used in this paper is slightly different from
those in \cite{op}.
For this, we will explicitly 
explain the correspondence.
First, for $\omega  \geq \Delta+l$, we can see that 
\begin{align}
\cO_{\Delta \omega  lm} = 
{ 2^{-(2n+l)} \over \int d \Omega}
{1 \over n!} { \Gamma({d \over 2} ) \over  \Gamma(n+{d \over 2}+l ) }
 s^{\mu_1 \mu_2 \ldots \mu_l}_{(l,m)}  P_{\mu_2} \cdots P_{\mu_l} 
(P^2)^{n} \hat{{\cal O}}_\Delta,
\label{rel2}
\end{align}
where 
$P^\mu$ act on an operator $\hat{\phi}$ such that
$P^\mu \hat{\phi} =[\hat{P}^\mu, \hat{\phi}]$, 
$\hat{{\cal O}}_\Delta 
= {1 \over  \int d \Omega} \cO_{\Delta \omega  lm}|_{\omega  =\Delta, l=m=0}$
and 
$n=(\omega  -\Delta-l)/2$ should be a non-negative integer. 
Here, $s^{\mu_1 \mu_2 \ldots \mu_l}_{(l,m)}$ is
a normalized rank $l$ symmetric traceless constant tensor which
is related to the normalized spherical harmonics by
\begin{align}
Y_{lm}(\Omega) = |x|^{-l} \,\,
s^{\mu_1 \mu_2 \ldots \mu_l}_{(l,m)} x_{\mu_1} x_{\mu_2} \cdots x_{\mu_l},
\end{align}
where they are assumed to be normalized as
\begin{align}
{1 \over \int d \Omega} \int d \Omega  \,\, Y_{lm}(\Omega) Y_{l' m' }(\Omega) 
=\delta_{l l'} \delta_{m m'},
\label{n1}
\end{align}
and $\int  d \Omega = {2 ( \pi)^{d/2} \over \Gamma(d/2)}$.
To obtain this, we repeatedly use 
\begin{align}
&(\partial_y^2)
\left(
(y^2)^n (s^{\mu_1 \mu_2 \ldots \mu_l}_{(l,m)} 
y^{\mu_1} y^{\mu_2} \cdots y^{\mu_l} )
\right) 
=4 n (n+l+d/2-1) (y^2)^{n-1} (s^{\mu_1 \mu_2 \ldots \mu_l}_{(l,m)} 
y^{\mu_1} y^{\mu_2} \cdots y^{\mu_l} ),
\end{align}
and, for $l \leq l'$,
\begin{align}
2 ^l {\Gamma\left( {l+d/2}\right) \over \Gamma\left( {d/2}\right)} \delta_{l l'} \delta_{m m'}
=
\,\,  
(s^{\nu_1 \nu_2 \ldots \nu_{l'}}_{(l',m')} \partial_{\nu_1} \partial_{\nu_2} \cdots \partial_{\nu_{l'}})
(s^{\mu_1 \mu_2 \ldots \mu_l}_{(l,m)} x_{\mu_1} x_{\mu_2} \cdots x_{\mu_l}).
\label{c1}
\end{align}
Then, the creation operator of the free theory in the naive large $N$ limit is given by
\begin{align}
\hat{a}_{n l m}^\dagger =
\tilde{c}_{nl} \, 
\cO_{\Delta \omega  lm}
\label{ao}
\end{align}
where $c_{nl}$ is the normalization constant given by
\begin{align}
\tilde{c}_{nl}=\sqrt{
 {\Gamma(\Delta) \over  \Gamma(\Delta+n+l) } 
{\Gamma(\Delta+1 -{d \over 2})  \over  \Gamma(\Delta+1-{d \over 2} +n)  }
n! {\Gamma(n+{d \over 2}+l ) \over  \Gamma({d \over 2} )}
\int  d \Omega
},
\end{align}
which satisfies $[\hat{a}_{n l m},\hat{a}_{n' l' m'}^\dagger]=\delta_{n n'} \delta_{l l'} \delta_{m m'}$ 
in the naive large $N$ limit.

\subsection{OPE and the commutator algebra}

In this subsection, we will consider the relation 
between the OPE and the commutator algebra.
It is well known that the commutator algebra of the conserved charges are given by OPE
using the deformation of the integration contour, like in $CFT_2$. 
However, this deformation technique can not be used 
for the operators which are not conserved charges, for example, the generic modes of the energy momentum tensor.

Here, we will explain how to derive, in principle, the algebra defined by 
the commutators of the operators $\cO_{\Delta   lm} (|x|)$ from
the OPE of the corresponding primary field $\cO_\Delta(x)$, which is assumed to be given as
\begin{align}
\cO_\Delta(x) \cO_\Delta(y) = {1 \over (x-y)^{2 \Delta}} + \cdots,
\label{so1}
\end{align}
where $\cdots$ includes the $1/N$ suppressed terms
which can be expanded by the primary fields \cite{SD} and
(assuming the parity invariance)
can be written as a sum of the following term:
\begin{align}
{1 \over (x-y)^{2 \Delta +l- \Delta'}} u^{\mu_1} u^{\mu_2} \cdots u^{\mu_l} 
\cO^{\Delta'}_{\mu_1 \mu_2, \cdots \mu_l} (y),
\label{ex0}
\end{align}
where $u^{\mu}={x^\mu-y^\mu }$ and 
$\cO^{\Delta'}_{\mu_1 \mu_2, \cdots \mu_l} (y)$ is the (not necessary primary) fields with 
spin $l$ and conformal weight $\Delta'$.

In the radial quantization, the commutator of the two fields at the equal ``time'' $|x|=1$ is given by
\begin{align}
[\cO_1(x), \cO_2(y)]_{|x|=|y|=1} &=
\lim_{\epsilon \rightarrow 0}
\left(
\cO_1(x)|_{|x|=1+\epsilon} \, \cO_2(y)|_{|y|=1} -\cO_2(y)|_{|y|=1} \,  \cO_1(x)|_{|x|=1-\epsilon} 
\right),
\CR
& =  \lim_{\epsilon \rightarrow 0}
\left(
\left( \cO_1(x) \, \cO_2(y) \right)|_{|x|=1+\epsilon, |y|=1} 
-\left( \cO_2(y) \,  \cO_1(x) \right)|_{|x|=1-\epsilon, |y|=1} 
\right),
\label{ec1}
\end{align}
where 
\begin{align}
\cO_1(x)=\frac{\partial^p }{\partial^p |x|}  \cO_\Delta(x), \,\,\,
\cO_2(y)=\frac{\partial^q }{\partial^q |y|}  \cO_\Delta(y).
\end{align}
We will use the OPE expansion for evaluating the r.h.s. of (\ref{ec1}).
One might worry about the other operator insertions near $x$ or $y$
which may invalidate the OPE expansion.
We will explain below this is not the case for the evaluation of the commutator.
Note that the equal time commutator of the local fields should vanish if $x \neq y$
because of the causality in the Lorentzian cylinder.
Thus, for non-vanishing commutator, $\left( \cO_1(x) \, \cO_2(y) \right)_{|x|=1 \pm \epsilon, |y|=1} $ 
in  (\ref{ec1}) can be 
given by evaluating  
$\frac{\partial^p }{\partial^p |x|}  \frac{\partial^q }{\partial^q |y|} 
\left( \cO_\Delta (x) \, \cO_\Delta (y) \right)$
with the OPE for $\cO_\Delta (x) \, \cO_\Delta (y) $
because $x=y$ and any other operator insertions are not close.
For $x \neq y$, the OPE is valid if there are no operator insertions near $x$ or $y$
and for this case the commutator should be zero using the OPE for the r.h.s. of (\ref{ec1}). 
However, even if there are operator insertions, the commutator should be zero 
because of the causality, although we can not use the OPE expansion.
Therefore, we can use the OPE for evaluating the r.h.s. of (\ref{ec1}) for any $x,y$.  

Furthermore, only the singular parts of the OPE can contribute the commutator
because ``regular'' terms vanishes if we take a limit $x \rightarrow y$.
(Here, the ``regular'' terms mean terms which vanish in the limit.
However, the terms with non-integer conformal weight will become singular term
by acting $\frac{\partial^p }{\partial^p |x|}$ with sufficiently large $p$.)

Because the commutator (\ref{ec1}) will contain a delta function for the ``space'' directions $\Omega$ as usual,
it will be convenient to consider the commutator of the modes of fields which are decomposed by the spherical Harmonics of ``space'' directions:  
\begin{align}
&\left[ \frac{\partial^p }{\partial^p |x|}  \cO_{\Delta  lm } (|x|) , \frac{\partial^q }{\partial^q |y|}  \cO_{\Delta  lm } (|y|) \right]_{|x|=|y|=1}  \CR
&={1 \over \left(\int  d \Omega \right)^2 }
\int d \Omega Y_{lm}(\Omega) \int d \Omega'  Y'_{l' m'}(\Omega') 
[\cO_1(x), \cO_2(y)]_{|x|=|y|=1}  ,
\label{ecm}
\end{align}
where $y$ also are decomposed to the radial direction $|y|$ and the angular directions $\Omega'$ for $S^{d-1}$.

We will first concentrate on the large $N$ leading term
in the OPE (\ref{so1}).
Then, we have the expansion of the correlation function as 
\begin{align}
{1 \over (x-y)^{2 \Delta}}=
{1 \over (r_{>})^{2 \Delta} }   { 1 \over \left( 1 + \left( {r_{<} \over  r_{>}} \right)^2- 2  \left({r_{<} \over  r_{>}}\right) \eta  \right)^\Delta }
=
{1 \over (r_{>})^{2 \Delta} } 
\sum_{s=0}^\infty \left( r_{<} \over r_{>} \right)^s C_s^\Delta \left(\eta \right) ,
\label{prop1}
\end{align}
where $\eta={x^\mu y_\mu \over |x| |y| } $, 
\begin{align}
C_s^\Delta \left(\eta \right) 
=\sum_{p=0}^{[\frac12 s]} 
\frac{(-1)^p (2 \eta)^{s-2p} }{p! (s-2p)!}
{\Gamma(\Delta+s-p) \over \Gamma(\Delta)},
\end{align}
is the Gegenbauer polynomial \cite{Avery} which reduces to the Legendre polynomial
for $\Delta=1/2$ and $r_{>}$ and $r_{<}$ are the larger and smaller ones of $|x|$ and $|y|$, respectively.
By the rotational invariance,
we can write 
\begin{align}
C_s^\Delta \left(\eta \right) 
= \sum_{n=0}^{[\frac12 s]} (d^{\Delta})_{s}^{\, s-2n} \, \sum_m Y_{(s-2n) m}^*(\Omega) Y_{(s-2n) m}(\Omega'),
\end{align}
where $\Omega$ and $\Omega'$ are the angular variables of $x$ and $y$,
respectively, and $(d^{\Delta})_{s}^{\, l} =0$ for $l>s$
or $s-l$ is odd.
For $\Delta=d/2-1$, we can see
$(d^{\Delta})_{s}^{\, l} = \frac{(d-2) 2 \pi^{d \over 2}}{(d+2 s-2) \Gamma(d/2)} \delta_{s l}$ 
\cite{Avery}.
This implies the relations between $\sum Y^*(\Omega)  Y(\Omega') $ and $(2\eta)^{s-2p}$:
\begin{align}
\frac{(d-2) 2 \pi^{d \over 2}}{(d+2 s-2) \Gamma(d/2)}
\sum_m Y_{(s-2n) \, m}^*(\Omega) Y_{(s-2n) \, m}(\Omega')
=\sum_{p=0}^{[\frac12 s]}  
\frac{(-1)^p (2 \eta)^{s-2p} }{p! (s-2p)!}
{\Gamma(d/2-1+s-p) \over \Gamma(d/2-1)}.
\end{align}
Using these relations, we find
\begin{align}
(d^{\Delta})_{s}^{\, s-2 n} = 
\frac{ 2 \pi^{d \over 2}}{ \Gamma(d/2)} 
{\Gamma(\Delta+s-n)  \over \Gamma(\Delta)   } 
{ \Gamma(\Delta+1-{d \over 2} +n)  \over \Gamma(\Delta+1 -{d \over 2})   } {1 \over n!} 
{ \Gamma({d \over 2} ) \over \Gamma(s-n+{d \over 2} ) },
\end{align}
and the correlator is expressed as 
\begin{align}
{1 \over (x-y)^{2 \Delta}}=
{1 \over (r_{>})^{2 \Delta} } 
\sum_{s=0}^\infty \left( r_{<} \over r_{>} \right)^s 
\sum_{n=0}^{[\frac12 s]} (d^{\Delta})_{s}^{\, s-2n} \, \sum_m Y_{(s-2n) m}^*(\Omega) Y_{(s-2n) m}(\Omega').
\label{prop2}
\end{align}

Thus, in the naive large $N$ limit, the equal time commutator between the modes is given by
\begin{align}
&\left[ \frac{\partial^p }{\partial^p |x|}  \cO_{\Delta  lm } (|x|) , \frac{\partial^q }{\partial^q |y|}  \cO_{\Delta  lm } (|y|) \right]_{|x|=|y|=1} 
\CR
&=
{1 \over \left(\int  d \Omega \right)^2 }
\sum_{s=0}^l \delta_{l l'}\delta_{m m'} \, (d^{\Delta})_{s}^{\, l} 
\left(
\left(  \frac{\partial^p }{\partial^p r} r^{-s-2 \Delta} 
\right)_{r=1} 
\left(
\frac{\partial^q }{\partial^q r} r^{s} 
\right)_{r=1}
-
\left(  \frac{\partial^q }{\partial^q r} r^{-s-2 \Delta} 
\right)_{r=1} 
\left(
\frac{\partial^p }{\partial^p r} r^{s} 
\right)_{r=1}
\right),
\label{ecm2}
\end{align}
where we have used
\begin{align}
&\lim_{\epsilon \rightarrow 0} \int d \Omega Y_{lm}(\Omega) \int d \Omega'  Y'_{l' m'}(\Omega') 
 \, 
\left[ 
\frac{\partial^p }{\partial^p |x|}  \frac{\partial^q }{\partial^q |y|} 
\left( {1 \over (x-y)^{2 \Delta}}  \right)
\right]_{|x|=1+\epsilon, |y|=1}  \CR
=&
\sum_{s=0}^l \delta_{l l'}\delta_{m m'} \, (d^{\Delta})_{s}^{\, l} 
\left(
\frac{\partial^p }{\partial^p r} r^{-s-2 \Delta} 
\right)_{r=1} 
\left(
\frac{\partial^q }{\partial^q r} r^{s} 
\right)_{r=1},
\end{align}
and similar one for $|x|=1-\epsilon$.
Note that the two different expansions, which originate from
the different limits of the integration contours, ($|x|=1+\epsilon$ and $|x|=1-\epsilon$), of 
the same function give the different results which make the commutator non-vanishing.
Note also that 
the commutator for the modes is indeed finite
because there are no infinite summations.

If we know the spectrum, we can also find 
the modes with fixed energy by the expansion: 
$\cO_{\Delta  lm } (|x|) = \sum_\omega |x|^{\omega-\Delta} \cO_{\Delta \omega lm }$.
Then, the commutator of the operators $\cO_{\Delta \omega lm }$ should satisfy (\ref{ecm2}).
Comparing the exponent of $|x|,|y|$,
a simple solution is 
\begin{align}
\left[\cO_{\Delta  \omega lm } ,  \cO_{\Delta  \omega' lm }  \right]
& = 
{1 \over \left(\int  d \Omega \right)^2 }
\sum_{s=0}^l \delta_{l l'}\delta_{m m'} \, (d^{\Delta})_{s}^{\, l} 
\left( 
\delta_{-s-2 \Delta  \,\,  \omega-\Delta} \delta_{s \,\, \omega'-\Delta}
-
\delta_{-s-2 \Delta  \,\,  \omega'-\Delta} \delta_{s \,\, \omega-\Delta}
\right), \CR
& =  
{1 \over \left(\int  d \Omega \right)^2 }
\delta_{\omega +\omega'}
\delta_{l l'}\delta_{m m'}
\, \left( 
(d^{\Delta})_{-\Delta+\omega'}^{\, l} 
-(d^{\Delta})_{-\Delta+\omega}^{\, l}
\right) 
\CR
& 
= 
{1 \over \left(\int  d \Omega \right)^2 }
\delta_{\omega +\omega'}
\delta_{l l'}\delta_{m m'} 
{\rm sgn} (\omega') (d^{\Delta})_{|\omega'|-\Delta}^{\, l},
\label{ecm3}
\end{align}
which agrees with (\ref{ao}).

Let us consider the full OPE, not only the large $N$ leading term, which includes terms written as
\begin{align}
{1 \over (x-y)^{2 \Delta +l- \Delta'}} u^{\mu_1} u^{\mu_2} \cdots u^{\mu_l} 
\cO^{\Delta'}_{\mu_1 \mu_2, \cdots \mu_l} (y).
\label{ex1}
\end{align}
Then, the ${1 \over (x-y)^{2 \Delta +l- \Delta'}} $ factor can be 
expanded by (\ref{prop2}) and 
$\cO^{\Delta'}_{\mu_1 \mu_2, \cdots \mu_l} (y)$ also can be expanded 
by the spherical functions and the symmetric tensor harmonics for $d>3$.
In order to compute the commutator, as for the leading order,
we just need to decompose the products of irreducible representations
of $SO(d)$. For $d=3$ the Clebsh-Gordon coefficients are well known 
and for other $d$ we can compute it, order by order, at least.
Thus, in principle, we can compute the commutator and 
it contains only a finite number of terms as for the large $N$ leading order.
It should be noted that
the terms (\ref{ex1}) 
which is non-divergent at $x-y=0$, 
do not contribute to the commutator 
because the two expansions are same.
Therefore, the commutators
of the modes,  
$\left[ \frac{\partial^p }{\partial^p |x|}  \cO_{\Delta  lm } (|x|) , \frac{\partial^q }{\partial^q |y|}  \cO_{\Delta  lm } (|y|) \right]_{|x|=|y|=1} $,
can be determined by the singular terms of the OPE,
which satisfies $2 \Delta  - \Delta' > 0$ if $2 \Delta  - \Delta'$ is integer, or
$2 \Delta  - \Delta' +p> 0$ if $2 \Delta  - \Delta'$ is not integer.

For the modes with fixed energy,
the commutator of the operators $\cO_{\Delta \omega lm }$ should satisfy (\ref{ecm}) 
and comparing the exponent of $|x|,|y|$,
a simple solution for the commutator of the operators with a general OPE is 
\begin{align}
[\cO_{\Delta \omega lm }, \cO_{\Delta \omega' l'm' }]
=& \int d \Omega Y_{lm}(\Omega) \int d \Omega'  Y'_{l' m'}(\Omega') 
\CR
& \, 
\left(
{\rm P}_+[|x|^{\Delta-\omega} |y|^{\Delta-\omega'} \cO_\Delta (x) \cO_\Delta (y)]
-{\rm P}_-[|x|^{\Delta-\omega} |y|^{\Delta-\omega'} \cO_\Delta (x) \cO_\Delta (y)]
\right),
\label{com1}
\end{align}
where ${\rm P}_{\pm} [f(x,y)]$ is defined as follows:
Let us consider a function $f(x,y)$ which has an expansion of the form
as $f(x,y)=|x|^a |y|^{-b} \sum_{m=0}^\infty (|y|/|x|)^{m} f_m(\Omega,\Omega')$
for $|x|>|y|$.
Then, ${\rm P}_{+} [f(x,y)]$ is the zero-mode of this expansion,
i.e. ${\rm P}_{+} [f(x,y)]=  f_a (\Omega,\Omega')$ if $a=b$ and $a$ is non-negative integer,
otherwise  ${\rm P}_{+} [f(x,y)]=0$.
On the other hand, for the function $f(x,y)$ which 
is written as $f(x,y)=|y|^{a'} |x|^{-b'} \sum_{m=0}^\infty (|x|/|y|)^{m} f_m(\Omega,\Omega')$ for $|y|>|x|$, 
${\rm P}_{-} [f(x,y)]=  f_a (\Omega,\Omega')$ if $a'=b'$ and $a'$ is non-negative integer,
otherwise  ${\rm P}_{-} [f(x,y)]=0$.
Note that the OPE of $\cO_\Delta(x) \cO_\Delta(y) $ has such expansions
depending on the sign of $|x|-|y|$.
We can check this formula for the leading order case.
With these expansion formulas, we can evaluate (\ref{com1}) as
\begin{align}
\int d \Omega Y_{lm}(\Omega) \int d \Omega'  Y'_{l' m'}(\Omega') 
 \, 
{\rm P}_+ \left[ |x|^{\Delta-\omega} |y|^{\Delta-\omega'} 
{1 \over (x-y)^{2 \Delta}} \right] 
= & \sum_{q=0}^\infty \delta_{\Delta+\omega+q } \delta_{\Delta-\omega'+q }
\delta_{l l'}\delta_{m m'}
\, (d^{\Delta})_{q}^{\, l} \CR
=& \delta_{\omega +\omega'}
\delta_{l l'}\delta_{m m'}
\, (d^{\Delta})_{-\Delta+\omega'}^{\, l},
\end{align}
which can be non-zero for $\Delta \leq \omega' \leq \Delta+l$, and
\begin{align}
\int d \Omega Y_{lm}(\Omega) \int d \Omega'  Y'_{l' m'}(\Omega') 
 \, 
{\rm P}_- \left[ |x|^{\Delta-\omega} |y|^{\Delta-\omega'} 
{1 \over (x-y)^{2 \Delta}} \right] 
= & \sum_{q=0}^\infty \delta_{\Delta+\omega'+q } \delta_{\Delta-\omega+q }
\delta_{l l'}\delta_{m m'}
\, (d^{\Delta})_{q}^{\, l} \CR
=&  \delta_{\omega +\omega'}
\delta_{l l'}\delta_{m m'}
\, (d^{\Delta})_{-\Delta+\omega}^{\, l},
\end{align}
which can be non-zero for $\Delta \leq \omega \leq \Delta+l$.
Thus, the commutators is given by 
\begin{align}
[\cO_{\Delta \omega lm }, \cO_{\Delta \omega' l'm' }]
= & \delta_{\omega +\omega'}
\delta_{l l'}\delta_{m m'}
\, \left( 
(d^{\Delta})_{-\Delta+\omega'}^{\, l} 
-(d^{\Delta})_{-\Delta+\omega}^{\, l}
\right) +\cdots
\CR
= &\delta_{\omega +\omega'}
\delta_{l l'}\delta_{m m'} 
{\rm sgn} (\omega') (d^{\Delta})_{|\omega'|-\Delta}^{\, l} +\cdots,
\end{align}
which indeed coincides with (\ref{ecm3}).

In this subsection, we have seen that 
with the singular parts of the OPE, 
the commutator of the operators are given 
by (\ref{ecm}) or (\ref{com1}) using the expansion
(\ref{prop1}), at least, in principle.
This will be used  below.

\subsection{Algebra for the  energy momentum tensor}

As for the scalar, we will consider the commutator algebra for the energy momentum tensor.

We will expand the energy momentum tensor 
with the traceless and conserved properties.
First, we will use the coordinates $r=|x|$ and $z^i$ as 
the coordinates of $S^{d-1}$, where $(i=1,\ldots,d-1)$,
with the flat metric $ds^2 = dr^2 +r^2 g^{S^{d-1}}_{ij} dz^i dz^j$.
We decompose it as 
\begin{align}
T_{\mu \nu} (x)  = 
 {\partial z^i  \over \partial x^\mu }  {\partial z^j  \over \partial x^\nu } T_{i j} (x) 
+ {x^\mu \over r}   {\partial z^j  \over \partial x^\nu } T_{r j} (x) 
+ {x^\nu \over r}   {\partial z^i  \over \partial x^\mu } T_{i r} (x) 
+ {x^\mu \over r}  {x^\nu \over r} T_{r r} (x) ,
\end{align}
where
\begin{align}
T_{i j} (x) & = e^\mu_i e^\nu_j T_{\mu \nu} (x) , \CR
T_{r j} (x) & = {x^\mu \over r} e^\nu_j T_{\mu \nu} (x) , \CR
T_{r r} (x) & = {x^\mu \over r}  {x_\nu \over r} T_{\mu \nu} (x) ,
\end{align}
and $e^\mu_i = {\partial x^\mu \over \partial z^i }$ $e^\mu_r = {\partial x^\mu \over \partial r }= {x^\mu \over r}$.
Then, for the $S^{d-1}$ directions, we will uniquely decompose them as
\begin{align}
T_{ij}(r,z^i)   =&  
\sum_{\omega
} r^{\omega-d+2}
\left(
\sum_{l=2}  \sum_{m} t^T_{\omega lm} \, Y^{lm}_{ij} (z^i)
 + \sum_{l=2, m} t^V_{\omega lm} \, (D_i Y_j^{lm} (z^i)+D_j Y_i^{lm} (z^i)) \right.
\nonumber \\ 
& \left. + \sum_{l=2} \sum_{m} t^S_{\omega lm} \, (D_i D_j - {1 \over d-1} g^{S^{d-1}}_{ij} D_k D_k )  Y^{lm} (z^i)
+\sum_{l=0} \sum_{m} t^{trace}_{\omega lm} {1 \over d-1} g^{S^{d-1}}_{ij} Y^{lm} (z^i)
\right),
\label{te}
\end{align}
\begin{align}
T_{r i}(r,z^i)  =
\sum_{\omega
} r^{\omega-d+1}
\left(
\sum_{l=1}  \sum_{m} v^V_{\omega lm} \, Y^{lm}_{i} (z^i) + \sum_{l=1} \sum_{m} v^S_{\omega lm} \, D_i Y^{lm} (z^i)
\right),
\end{align}
and
\begin{align}
T_{r r}(r,z^i)  =
\sum_{\omega
} r^{\omega-d}
\sum_{l=0} \sum_{m} s^S_{\omega lm} \,  Y^{lm} (z^i).
\end{align}
Here,
the rank $r$ symmetric (traceless)
tensor harmonics on unit radius $S^{d-1}$, $Y^{lm}_{i_1,i_2, \cdots, i_r}(\Omega)$,
is defined such that
$Y^{lm}_{i_1,i_2, \cdots, i_r}$ is totally symmetric for the indices $i_k$
and
\begin{align}
& D^i D_i Y^{lm}_{j_1,j_2, \cdots, j_r}
=(-l(l+d-2)+r) Y^{lm}_{j_1,j_2, \cdots, j_r}, \\
& D^i Y^{lm}_{i,i_2, \cdots, i_r}=0 \\
& g^{ij}_{S^{d-1}} Y^{lm}_{i,j,i_3 \cdots, i_r}=0,
\end{align}
where
$z^i$ $(i=1,2, \cdots,d-1)$ are coordinate of $S^{d-1}$, 
$D_i$ is the covariant derivative on $S^{d-1}$,
and $g^{ij}_{S^{d-1}}$ is the inverse metric of unit radius $S^{d-1}$. 
Here, 
$l=r,r+1,r+2,\cdots$ and
$m$ runs from $1$ to the number of the independent 
harmonics which depends on $l$ and $r$.
This harmonics $Y^{(r)lm}_{i_1,i_2, \cdots, i_r}$
is the unitary representation 
of $SO(d)$
which corresponds to the Young diagram labeled
by $[l,r,0,\ldots,0]$.
More details for the symmetric tensor harmonics, see \cite{SSH1,SSH2,SSH3}.

The energy $\omega$ should be an integer in the naive large $N$ limit
because there are no states with non-integer energy
for the primary field with the integer conformal weight.
This implies that a possible energy $\omega$ is an integer except $1/N^2$ corrections,
which is expected from the usual large $N$ expansion.
Note that the modes with fixed energy, for example $t^T_{\omega l m}$, may be
identically zero depending on $\omega$.

We will show that
the rank $r$ symmetric (traceless)
tensor harmonics can be represented by
\begin{align}
Y^{lm}_{j_1,j_2, \cdots, j_r}=
s^{rlm}_{\mu_1 \mu_2 \cdots \mu_r ; \nu_1 \nu_2 \cdots \nu_{l} }
{e^{\mu_1}_{j_1}  \over r} {e^{\mu_1}_{j_2}  \over r} \cdots {e^{\mu_1}_{j_r}  \over r} \,
{x^{\nu_1} \over r} {x^{\nu_2} \over r} \cdots {x^{\nu_{l}} \over r},
\end{align}
where $l \geq r $ and
$s^{rlm}_{\mu_1 \mu_2 \cdots \mu_r ; \nu_1 \nu_2 \cdots \nu_l}$
is a traceless constant tensor
which is
given by the anti-symmetrization of 
the $r$ pairs of the following indices:
$(\mu_1, \nu_1), (\mu_2, \nu_2), \cdots,  (\mu_r, \nu_r)$,
and then the symmetrization for the $\mu_a$ and $\nu_a$.
This (anti)symmetrization procedure corresponds to the Young diagram labeled
by $[l,r,0,\ldots,0]$.
Using 
\begin{align}
\delta^{\mu}_{ \nu }=\pa_\rho x^\mu \pa^\rho x^\nu=
\nabla_i  x^\mu \nabla^i  x^\nu +\nabla_r  x^\mu \nabla^r  x^\nu  = 
\frac{1}{r^2}  g^{ij}_{S^{d-1}} e_i^\mu e_j^\nu +\frac{1}{r^2} x^\mu x^\nu,
\end{align}
where $\nabla$ is the covariant derivative on $\R^{d}$,
we can see that 
$ g^{ij}_{S^{d-1}} Y^{lm}_{i,j,i_3 \cdots, i_r}=0$ because $s^{rlm}_{\mu_1 \mu_2 \cdots \mu_r ; \nu_1 \nu_2 \cdots \nu_{l-r} }$ 
is traceless and anti-symmetric for  the interchanging a pair of $\mu$ and $\nu$.
Next, note that the Christoffel symbols in the coordinates $\{ z^i,r \}$ of $\R^{d}$
are given by
\begin{align}
\Gamma^i_{jr} =\frac{1}{r} \delta^i_j= \Gamma^i_{rj}, \,\,\,\,
\Gamma^r_{ij}=-r g_{ij}^{S^{d-1}},  \,\,\,\,  \Gamma^k_{ij}=\Gamma^{S^{d-1} \,\, k  }_{ \,\,\,\,\, ij},
\end{align}
where $\Gamma^{S^{d-1} \,\, k  }_{ \,\,\,\,\, ij}$ is the Christoffel symbols of the $S^{d-1}$,
and others vanish.
Then, we can show that
\begin{align}
0 & =e^\alpha_i e^\beta_j \nabla_\alpha \nabla_\beta x^\mu
=\nabla_i  \nabla_j x^\mu 
=
D_i e^\mu_j + g_{ij}^{S^{d-1}} x^\mu,
\end{align}
which implies $  D^i e^\mu_i =- (d-1) x^\mu $, $D^i D_i e^\mu_j =-  e_j^\mu$ and
$g^{jk}_{S^{d-1}} e_j^\nu D_i e^\mu_k =- e_i^\nu x^\mu$.
Using these, we can see that 
$ D^i Y^{lm}_{i,i_2, \cdots, i_r}=0 $ and 
$D^i D_i Y^{lm}_{j_1,j_2, \cdots, j_r}
=(-l(l+d-2)+r) Y^{lm}_{j_1,j_2, \cdots, j_r}$
hold because of the 
traceless and (anti)symmetric properties of $s^{nlm}$.

In the expansion (\ref{te}), the traceless condition of the energy momentum tensor 
is just 
\begin{align}
t^{trace}_{\omega lm}=-s^S_{\omega lm}.
\end{align}
For the conservation condition $\partial^\mu T_{\mu \nu}=0$, we will use following formula:
\begin{align}
\partial^\mu T_{\mu r} &=
\left( {\partial \over \partial r } + \frac{d}{r}
\right) T_{rr} + \frac{1}{r^2} D^i T_{ir}, \CR
\partial^\mu T_{\mu i} &=
\left( {\partial \over \partial r } +\frac{d-1}{r}
\right) T_{ri} + \frac{1}{r^2} D^j T_{ji}.
\end{align}
With these, we find, for $l \ge 2 $,
\begin{align}
\omega s^S_{\omega lm} = l(l+d-2)  v^S_{\omega lm},  \,\,\,\, 
\omega  v^V_{\omega lm} = (l+d-1)(l-1) t^V_{\omega lm} \\
s^S_{\omega lm} =\omega  (d-1) v^S_{\omega lm} - (d-2 )  (l+d-1)(l-1) t^S_{\omega lm},
\end{align}
where we have used $(D_i D_j-D_j D_i) V_k=V_l R^l_{\,\, ikj}$ and $R_{ijkl}=g_{il}g_{jk}-g_{ik}g_{jl}$
for unit radius sphere.
Thus, for $l \ge 2 $, only the $s^S_{\omega lm} , v^V_{\omega lm}, t^T_{\omega lm} $
are the independent operators for the energy momentum tensor.
We find, for $l=1$,  
\begin{align}
\omega  v^V_{\omega lm} =0, \,\,\,
\omega s^S_{\omega lm} = (d-1)  v^S_{\omega lm},  \,\,\,\, 
s^S_{\omega lm} =\omega  (d-1) v^S_{\omega lm} ,
\end{align}
with $ t^S_{\omega lm} =t^V_{\omega lm}=t^T_{\omega lm}=0$,
and, for $l=0$,  
\begin{align}
\omega s^S_{\omega lm} = 0,  \,\,\,\, 
\end{align}
with $ t^S_{\omega lm} =t^V_{\omega lm}=t^T_{\omega lm}=v^V_{\omega lm} =v^S_{\omega lm} =0$.
Thus, the non-trivial operators for $l=0,1$ are
$v^V_{0 \,  1 \, m}, s^S_{\pm 1 \,  1 \, m}=v^S_{\pm 1 \,  1 \, m}$ and $s^S_{0 \,  0 \, 0}$.
All of these correspond to the generators of the conformal symmetry.
Indeed, inserting the expansion of the $T_{\mu \nu}(x)$ into the definition of the generators,
\begin{align}
Q_{\epsilon}(S^{d-1})= - \int_{S^{d-1}} d S_{\mu} \, \epsilon_\nu \, T^{\mu \nu}(x),
\end{align}
where $ \epsilon_\nu$ is the conformal Killing vector and integration is on $S^{d-1}$ at a fixed $r$,
we find that
$D \sim s^S_{0 \, 0 \, 0}, M_{\mu \nu} \sim s^{11m}_{\mu \nu}  v^V_{0 \,  1 \, m}, P^\mu \sim s^{01m}_{\mu} s^S_{1 \,  1 \, m}, K^\mu \sim s^{01m}_{\mu} s^S_{-1 \,  1 \, m}$
for any choice of $r$ as required from the conservation law.
Note that all the generators of the conformal symmetry have $l=0,1$ in our notation.

We will denote a set of the non-trivial $s^S_{\omega lm} , v^V_{\omega lm}, t^T_{\omega lm} $ as 
\begin{align}
L_{A \omega lm} \equiv \{ s^S_{\omega lm} , v^V_{\omega lm}, t^T_{\omega lm} \}, 
\end{align}
where the index $A$ takes $S,V,T$. 
We define also the mode of the energy momentum tensor
decomposed for the space direction $\Omega$ as
\begin{align}
L_{A lm} (|x|) = \sum_\omega r^{\omega-d-r[A]+2} L_{A \omega lm}.
\end{align}

Let us consider the spectrum of the low energy theory.
The energy momentum tensor is Hermite on the cylinder, then,
\begin{align}
L_{A \omega lm}^\dagger =  L_{A (-\omega )l m},
\end{align}
where we have taken $Y^{lm}_{i_1, i_2, \cdots, i_r}$ real.
Here, the energy momentum tensor on the cylinder ($r=e^\tau$) is given by 
\begin{align}
T_{ij}^{cy} (\tau,z^i)   =  r^{d-2} T_{ij}(r,z^i) , \,\,
T_{\tau i}^{cy} (\tau,z^i)   =  r^{d-1} T_{r i}(r,z^i), \,\,
T_{\tau i}^{cy} (\tau, \tau)   =  r^{d} T_{r r}(r,z^i),
\end{align}
where extra $1/r$ factors are from the normalizing $dx^{\mu}/d z^i$.\footnote{
We ignored the Weyl anomaly here because it is a constant and does not play
important roles in this paper. 
}
This energy momentum tensor indeed satisfies the conservation law.

As for the scalar case, we will require the regularity 
of $T_{\mu \nu}(x) \ket{0}$ at $r=0$.
For $L_{T \omega lm}$, this means 
$L_{T \omega lm} \ket{0}=0$ for $\omega <d+l $
because 
$ {\partial z^i  \over \partial x^\mu }  {\partial z^j  \over \partial x^\nu } 
Y^{lm}_{ij} (z^i)
=  s^{rlm}_{\mu_1 \mu_2 ; \nu_1 \nu_2 \cdots \nu_{l} }
\delta^{\mu_1}_\mu \delta^{\mu_2}_\nu \frac{1}{r^2}
\,
{x^{\nu_1} \over r} {x^{\nu_2} \over r} \cdots {x^{\nu_{l}} \over r}$.
Similarly, 
we can see that 
\begin{align}
L_{A \omega lm} \ket{0}=0 \mbox{ for } \omega <  d+l+r[A]-2,
\label{cond1}
\end{align}
where $r[A]=0,1,2$ for $A=S,V,T$, respectively.
This can be checked by considering $T_{\mu \nu}(x)=c_{\mu \nu}$, which is constant and traceless.  
This is the lowest regular term and in the polar coordinates,
$T_{i j} (x) = e^\mu_i e^\nu_j c_{\mu \nu}, \,\,
T_{r j} (x)  = {x^\mu \over r} e^\nu_j c_{\mu \nu}, \,\,
T_{r r} (x) = {x^\mu \over r}  {x_\nu \over r} c_{\mu \nu}$.
This corresponds to $s^S_{\omega lm}$ with $\omega=d, l=2$,
which is the boundary of the condition (\ref{cond1}). 
For $L_{V \omega lm}$, we can check (\ref{cond1}) by
considering $T_{\mu \nu}(x)=c_{\mu \nu \rho} x^\rho$.

The Hermite condition implies that  $\bra{0}  L_{A \omega lm}=0$ for $\omega  > -( d+l+r[A]-2)$.
Thus the operators $L_{A \omega lm}$ 
with $|\omega | < d+l+r[A]-2$ satisfy 
$0=L_{A \omega lm} \ket{0} =\bra{0}  L_{A \omega lm}$
and we will denote these operators 
as $L^{iso}$, which includes the generators of the 
conformal group.
We will also denote the operators $L_{A \omega lm} $ 
with $\omega   \geq d+l+r[A]-2$ and $\omega   \leq -(d+l+r[A]-2)$ 
as $L^{+}$  and  $L^{-}$, respectively.

In the naive large $N$ limit,
$\omega$ is integer and 
$\omega$ should be
restricted to satisfy $\omega -( d+l+r[A]) \in 2 \z$
for $L^{\pm}_{\omega lm} $.
This restriction comes from the fact that 
the spectrum are constructed 
by acting $P^\mu$ on the primary states. 
It is important to note that $L^{+}_{\omega lm} $ correspond to
the creation operators of the free gravity theory in
the asymptotic $AdS_{d+1}$ \cite{IW}
in the naive large $N$ limit \cite{op}.

For a generic strong coupling large $N$ gauge theories
with conformal symmetry, the low energy primary field 
is energy momentum tensor only
and the spectrum generated by the it are expected to be independent as assumed in \cite{op}.
Thus, in the naive large $N$ limit, 
the (low energy) states are spanned by 
\begin{align}
|  {\cal N}_{A n lm}   \rangle
\equiv
\prod_{A,n \in \z_{\ge 0} ,l,m} 
{
(L^+_{A \omega lm})^{{\cal N}_{A \omega lm}} 
\over ({\cal N}_{A n lm})!
}
| 0 \rangle,
\label{state}
\end{align}
where ${{\cal N}_{A n lm}}$ is a non-negative integer,
$\omega= d+l+r[A]-2+2 n$ and 
\begin{align}
\hat{H} (\prod_{A,\omega,l,m} (L^+_{A \omega lm})^{{\cal N}_{A n lm}}) | 0 \rangle
=(\sum_{A, n, l, m} {{\cal N}_{A n lm}} \, \omega) (\prod_{A,n,l,m} (L^+_{A \omega lm})^{{\cal N}_{A n lm}}) | 0 \rangle.
\end{align}
These coincide the states of the Fock space 
in the weak coupling limit of the gravity on AdS space as shown in \cite{op}.
These states are independent by the assumption in the limit taken in \cite{op},
however, will not be independent if we consider high-energy states for
a large, but, finite $N$ case.

Including the $1/N$ corrections, 
the conformal dimension of the energy momentum tensor 
is not modified, however, the dimension of the primary fields of multi trace operators will be modified slightly.
Because this modification is small,
the (low energy) states will be still spanned by 
energy eigen states labeled by $\{ {\cal N}_{A n lm} \} $:
\begin{align}
|  {\cal N}_{A n lm}  \rangle,
\end{align}
where we used same notation for the states 
in the naive large $N$ limit and their deformations 
by the $1/N$ corrections.
These eigen states $|  {\cal N}_{A n lm}  \rangle$
are also be generated by the primary (multi trace) fields and 
their descendants.
This means that in the naive large $N$ limit
the eigen state $|  {\cal N}_{A n lm}  \rangle$ reduces 
to the r.h.s. of (\ref{state}).
Note that the state
$| {\cal N}_{A n lm} \rangle$ can not be 
simply given by (\ref{state}) using an analogue of $L^+_{A \omega lm}$ and 
the energy of $|  {\cal N}_{A n lm}   \rangle$ is no longer need to be an integer.

We will also define an deformation of 
$L_{A \omega lm}$ of the naive large $N$ limit to 
the case including the $1/N$ corrections
by
\begin{align}
\bar{L}_{A n lm} \equiv {1\over 2 \pi} \int_{-\pi}^{\pi} d \beta
e^{-i (l+2n) \beta}  \,
\sum_{\omega} ( e^{i \beta})^{\omega-d-r[A]+2}
L_{A \omega lm},
\end{align}
where $n \in \z /2$
and $ L_{A lm} (e^{i \beta} |x|)|_{|x|=1} =\sum_\omega ( e^{i \beta})^{\omega-d-r[A]+2}
L_{A \omega lm}$ is 
the mode of 
the ``time'' translated energy momentum tensor $ \exp (i \beta |x| \frac{\de}{\de |x|}) T_{\mu \nu}(x) |_{|x|=1}$.
In the cylinder coordinate, not in the radial quantization of the flat space,
this indeed correspond to (formal) Fourier coefficient of the time translated energy momentum tensor,
although the field is not periodic in time including $1/N$ corrections.\footnote{
This definition of a deformation of $L_{A \omega lm}$ 
will satisfy non-trivial commutators with the $1/N$ corrections and 
reduce to the $L_{A \omega lm}$ in the large $N$ limit,
although there are other definitions satisfying these properties.
}
Because this is the ``Fourier transformation'', we can see that 
$\bar{L}_{A n lm}  = \sum_{\omega \sim (d+l+r[A]-2+2n) } L_{A \omega lm}
+{\cal O}(1/N^2)$
where $\omega \sim (d+l+r[A]-2+2n)$ means that 
$\omega-(d+l+r[A]-2+2n)={\cal O}(1/N^2)$,
i.e. $\omega=d+l+r[A]-2+2n$ in the naive large $N$ limit.
Note that $\bar{L}_{A n lm}$ is not an eigen operator of 
the Hamiltonian $D$ and the violation of that is ${\cal O}(1/N^2)$. The creation (and annihilation) operators in the naive large $N$ limit corresponds to the $\bar{L}_{A n lm}$ with $l \ge 2$
and $n \in \z_{\ge 0}$ (and $n+d+l+r[A]-2 \in \z_{\le 0}$), respectively.

Before considering the commutation relation, 
we first investigate the singular parts of 
the OPE of the energy momentum tensors.
For the  energy momentum tensor,
two point function is given by
\begin{align}
\langle T^{\mu_1 \nu_1}(x)  T^{\mu_2  \nu_2} (0) \rangle
=C_T \left(
{ 
I^{ \mu_1}_{\,\,\,\,\, \alpha} (x) 
I^{ \nu_l} _{\,\, \beta}  (x) 
\left(
\frac12 (\delta_{\alpha \mu_2} \delta_{\beta \nu_2}+\delta_{\alpha \nu_2} \delta_{\beta \mu_2})
+\frac1d \delta_{\alpha \beta} \delta_{\mu_2 \nu_2}
\right)
\over x^{2 d} } 
\right),
\end{align}
where $I^\mu_{\,\, \nu} (x)=\delta^\mu_{\,\,\, \nu}-2 {x^\mu x_\nu \over x^2 }$
and $C_T$ is a constant.
We use the usual normalization for the energy momentum tensor 
in which the conformal generators
are given by the mode expansions of the energy momentum tensor,
like the Virasoro algebra. 
This means $C_T ={\cal O}(N^2)$ for the large $N$ gauge theories.
The three point function is also fixed in \cite{Osborn} 
with the three coefficients ${\mathcal A,B,C}$ with which 
$C_T$ is written as
\begin{align}
C_T={I(0) \over 2} {(d-1)(d+2){\mathcal A}-2{\mathcal B}-4(d+1){\mathcal C} \over d(d+2)}.
\end{align}
Note that 
the singular parts of 
the OPE between the energy momentum tensors are
almost fixed, 
if the two and the three point functions are given,
at least in principle,  by the general argument of the CFT \cite{Qu, Ry, SD}.
This is because the primary fields which can appear in the singular part 
are only the identity operator, the energy momentum tensor
and the double trace operators, which correspond to composites of the two energy momentum tensors,
if the anomalous dimensions of them are negative.
We will neglect the multi trace operators for a while and consider them later.
Thus, the singular part of the OPE is also given with only three unknown coefficients  $A,B,C$
as
\begin{align}
T^{\mu_1 \nu_1}(x)  T^{\mu_2  \nu_2} (0) 
= & C_T \left(
{ 
I^{ \mu_1}_{\,\,\,\,\, \alpha} (x) 
I^{ \nu_l} _{\,\, \beta}  (x) 
\left(
\frac12 (\delta_{\alpha \mu_2} \delta_{\beta \nu_2}+\delta_{\alpha \nu_2} \delta_{\beta \mu_2})
+\frac1d \delta_{\alpha \beta} \delta_{\mu_2 \nu_2}
\right)
\over x^{2 d} } 
\right) \CR
&
+ s_{\mu_1 \nu_1 \mu_2 \nu_2 \mu_3 \nu_3}(x, \partial) T^{\mu_3 \nu_3}(0)+\cdots,
\label{opet}
\end{align}
where 
$ s_{\mu_1 \nu_1 \mu_2 \nu_2 \mu_3 \nu_3}(x, \partial)$
consists of terms proportional to 
${\mathcal A}/C_T$, ${\mathcal B}/C_T$ and  ${\mathcal C}/C_T$.
Because of the large $N$ factorization, we find
${\mathcal A}={\cal O}(N^2)$,
${\mathcal B}={\cal O}(N^2)$ and 
${\mathcal C}={\cal O}(N^2)$.\footnote{
We assumed that there are no low energy fields other than the energy momentum tensor.
}

As for the scalar case, 
the commutation relations between the 
the modes of the energy momentum tensor
with the ``time'' derivatives
$\frac{\de^p}{\de |x|^p} L_{A lm} (|x|)$ 
are fixed by the singular part of the OPE of the energy momentum tensor with the derivatives
although we will not calculate them because of the technical difficulties.
These commutation relations can be translated to 
the commutators betweens $\bar{L}_{A n lm}$
ordr by order, in principle.
Thus, neglecting the contributions of the multi-trace operator,
the commutators are given as
\begin{align}
[L_i, L_j]= N^2 \omega_{ij} {\mathbf 1} + f_{ij}^k L_k,
\label{alg}
\end{align}
where 
$\omega_{ij}$ and $ f_{ij}^k$ are $N$-independent constants and
we denoted $\bar{L}_{A n lm}$ (or $\frac{\de^p}{\de |x|^p} L_{A lm} (|x|)|_{|x|=1}$ ) as $L_i$, for short.
These constants are, in principle, fixed by the OPE (\ref{opet}) which is fixed by the two and
three point functions given in \cite{Osborn} which has only two parameters other than $N$.
This algebra (\ref{alg}) forms an infinite dimensional Lie algebra including the identity operator ${\mathbf 1}$
like the Virasoro algebra.

In the naive large $N$ limit, for the space of states (\ref{state}), which are valid in the low energy approximation,
any opeartor
can be represented as
a (formal) sum of the polynomials of 
the operators 
in $L^+$ and $L^-$. 
Indeed, for a state $\ket{{\cal N}_{A \omega lm}} = (\prod_{A,n,l,m} (L^+_{A \omega lm})^{{\cal N}_{A n lm}}) \ket{0}$ with the energy $E=\sum_{A, n, l, m} {{\cal N}_{A n lm}} \, \omega$,
we can show $(\prod_{A,n,l,m} (L^+_{A \omega lm})^{{\cal N}'_{A n lm}})^\dagger \ket{{\cal N}_{A n lm}}=0$
if $E'=\sum_{A, n, l, m} {{\cal N}'_{A n lm}} \, \omega > E$
because the vacuum is the lowest energy state.
Furthermore, the states $\ket{{\cal N}_{A n lm}}$ 
which have a same energy are independent
by the assumption.
Thus, we can construct an operator $L$ which
has the matrix element
$\bra{{\cal N}_{A n lm}} L \ket{{\cal N}_{A n lm}}$ from
the polynomials of $L^+$ and $L^-$
order by order according to the energy of the ket.
By replacing $L^+,L^-$ to the corresponding operators in $\bar{L}_{A n lm}$ with non-negative integer $n$,
above statements are clearly valid 
including the $1/N$ corrections,

\section{Classical limit of the $CFT_d$}

In this section, we consider what is the classical limit 
of the large $N$ $CFT_d$.
For this, we will first introduce two more different normalizations of $L_i$.
First, we define $L_i^F= \frac1N L_i$ which satisfy 
\begin{align}
[L^F_i, L^F_j]=  \omega_{ij} {\mathbf 1} + \frac1N f_{ij}^k L^F_k.
\label{algF}
\end{align}
This normalization corresponds to the usual normalization 
of the primary field other than conserved currents,
up to an ${\cal O}(N^0)$ factor,
while $L_i$ include the conformal generators with an
${\cal O}(N^0)$ factor.
This normalization is suitable especially for the free limit
where we neglect the last term in (\ref{algF}), then 
they becomes the creation and annihilation operators (and $L^{iso}$) on the Fock space.
The other normalization is defined by $L_i^{cl}= \frac{1}{N^2} L_i$ which satisfy 
\begin{align}
N^2 [L^{cl}_i, L^{cl}_j]=  \omega_{ij} {\mathbf 1} + f_{ij}^k L^{cl}_k.
\label{algc}
\end{align}
Note that the r.h.s. of this is $N$-independent.
If we regard $N^2$ as $1/\hbar$, 
it is expected, for $N \gg 1$,  that 
\begin{align}
-i \{ L^{cl}_i, L^{cl}_j \}_P=  \omega_{ij} {\mathbf 1} + f_{ij}^k L^{cl}_k,
\label{Poi}
\end{align}
where $\{ L^{cl}_i, L^{cl}_j \}_P$ is a Poisson bracket
of the corresponding classical theory where 
$L^{cl}_i $ is identified as $\bra{\Psi} L^{cl}_i \ket{\Psi}$
for a ``classical'' state $\ket{\Psi}$.
We will see this below.

First, let us explain how to obtain the 
classical limit of the quantum mechanics,
for example, for a particle.
Note that a quantum mechanical system need not 
to have a classical limit if the theory is not obtained from 
a quantization of a classical system. 
Indeed, there exist purely quantum mechanical systems.
On the other hand, 
let us consider a quantum mechanical system
where we can choose operators $\hat{p},\hat{x}$
with the commutation relations
$-\frac{i}{\epsilon} [\hat{x}^i, \hat{p}_j]=  \delta_{j}^i
+{\cal O}(\epsilon)$, where 
$\epsilon$  is a small parameter, and
the Hamiltonian 
$\hat{H}=\frac{1}{\epsilon} \hat{H}'(\hat{p},\hat{x})+{\cal O}(\epsilon^0)$.
Here, the time evolution of a operator $\hat{O}$ is assumed to be given by
$\frac{d \hat{O}}{dt}=i [\hat{H}, \hat{O}]$.\footnote{
If we regard $\epsilon$ as $\hbar$,
the usual definition of the Hamiltonian $\hat{H}'(=\epsilon \hat{H})$ satisfies
$\epsilon \frac{d \hat{O}}{dt}=i [\hat{H}', \hat{O}]$.
Thus, our definition of the Hamitonian has
a different normalization from the usual one.
This is because we would like to consider the CFT
where there is no notion of $\hbar$ generically.
}
Then,  regarding $\epsilon$ as $\hbar$,
this quantum system may be derived by the quantization from the classical system with 
the Poisson bracket $ \{ x^i, p_j \}_P= \delta_{j}^i$
and the Hamiltonian $H^{cl}=  \hat{H}'$
which satisfies $\frac{d O}{dt}=- \{ H^{cl}, O \}_P$.
For this identification with the classical system, \footnote{
Note that if we start from the quantum mechanics,
instead from the classical mechanics,
the classical limit and 
the parameter $\hbar$ emerge
if the theory satisfies some requirements.
In particular we need a small parameter,
which for our case $1/N^2 \sim 1/C_T$.
}
we need to consider a state with 
$x^i={\cal O}(\epsilon^0)$ and 
$p_j={\cal O}(\epsilon^0)$, which implies 
$H^{cl} ={\cal O}(\epsilon^0)$,
and then this state will have large ``quantum numbers''
because $\epsilon$ is small.

This classical limit of the quantum system is an approximation,
where states which have same $x^i, p_j$ up to ${\cal O(\epsilon)}$ differences are identified.
For the operators, we need to consider the ``classical operators'' ${\cal O}(x^i, p_j)$ 
which are also defined up to ${\cal O(\epsilon)}$ differences.
For the CFT case, $ L^{cl}_i$ indeed correspond to 
$\hat{x},\hat{p}$ above and give the classical limit with $\epsilon \sim 1/N^2$.\footnote{
More precisely, 
$\{  \frac{1}{N^2} L^{+ \, \, cl}_i \}$ or $ \{  \frac{\de^p}{\de r^p} T_{\mu \nu}(x)|_{r=1} \}$
generates any operator in the low energy states, thus 
they correspond to $x^i, p_j$.
}
Note that if we use $z^a(\hat{x},\hat{p})$ as basis of operators 
instead of $\hat{x},\hat{p}$,
$\frac{i}{\epsilon} [\hat{x}^i, \hat{p}_j]=-  \delta_{j}^i
+{\cal O}(\epsilon)$ will be replaced by
$\frac{i}{\epsilon} [z^a, z^b]=-  f^{ab}(z)
+{\cal O}(\epsilon)$ 
and 
these should satisfy the Jacobi identities.
The commutation relations (\ref{algc}) are 
in this generalized form, locally.

Let us consider the contributions of the multi trace operators.
The primary multi trace operators appear in the OPE between the energy momentum tensors, however, with $|x-y|^{m+\delta}$
where $m$ is non-negative integer and $\delta={\cal O}(1/N^2)$ is the anomalous dimension of the multi trace operator.
Thus, the contributions to the commutator is suppressed
by $\delta \sim 1/N^2$. Indeed, if $\delta=0$ then this term is regular
which can not contribute to the commutator.
Next we will consider how much such the term with a multi trace operator in the OPE is 
suppressed in the large $N$ limit.
First, we will consider scalar operators in the naive large $N$ limit
for simplicity because essentially same considerations can be applied to the energy momentum case. 
The OPE is exactly given by 
\begin{align}
\cO_{\Delta_a} (x) \cO_{\Delta_b} (y) = {1 \over (x-y)^{2 \Delta}} \delta_{ab}+
: (e^{z^\mu \frac{\de}{\de y^\mu}}  \cO_{\Delta_a}(y) )|_{z=x-y}   \cO_{\Delta_b}(y) :,
\label{so2}
\end{align}
where $: \cdots :$ is the free theory normal ordering,
because it is a generalized free theory.
Note that the $e^{z^\mu \frac{\de}{\de y^\mu}}$ generates just 
the descendants of $\cO_{\Delta_a}$. 
The classical state for this is the coherent state 
$\ket{\alpha}$ for which 
the expectation value is
$\bra{\alpha} \cO_{\Delta_a} (x)  \ket{\alpha}=\alpha_a (x)$.
Then, for the naive large $N$ limit where $\alpha_a (x) = \epsilon {\cal O} (N) \gg 1$,  with $\epsilon \ll 1$,
\begin{align}
\bra{\alpha} \cO_{\Delta_a} (x) \cO_{\Delta_b} (y) \ket{\alpha} 
& = {1 \over (x-y)^{2 \Delta}} \delta_{ab}+
\bra{\alpha} 
: (e^{z^\mu \frac{\de}{\de y^\mu}}  \cO_{\Delta_a} (y) )|_{z=x-y}   \cO_{\Delta_a} (y) : \ket{\alpha} \CR
&
\sim  (e^{z^\mu \frac{\de}{\de y^\mu}} \alpha_a (y) )|_{z=x-y}   \, \alpha_b (y) = \alpha_a (x) \alpha_b(y),
\label{vev2p}
\end{align}
which is ${\cal O} (N^2) $ and the correct results in the classical limit.
This implies that
double trace operator terms appear in the OPE, $\cO_{\Delta_a} (x) \cO_{\Delta_b} (y)$, 
including $1/N$ corrections are suppressed by $1/N$, at least,
except $[\cO_{\Delta_a} \cO_{\Delta_b}] (x)$.
For the multi trace operator $[ \prod_{m=1}^{p} \cO_{\Delta_{a_m}}] (x)$
in the OPE, 
this will gives ${\cal O}(N^p)$ contributions to $\bra{\alpha} \cO_{\Delta_a} (x) \cO_{\Delta_b} (y) \ket{\alpha} $
where $\ket{\alpha}$ is a classical state.
Such term is suppressed by $1/N^{p-1}$, at least,
for the consistency of the classical picture.\footnote{
We expect that this suppression is explained by the large $N$ factorization.}
Thus, including the suppression by the anomalous dimension $\delta$,
in the classical limit, the multi trace operator
contributions to the commutator is only 
the double trace operator 
$[\cO_{\Delta_a} \cO_{\Delta_b}] (x)$ with an unknown parameter $\delta$.
Therefore, the commtator including the multi trace operator contribution is
\begin{align}
N^2 [L^{cl}_i, L^{cl}_j]=  \omega_{ij} {\mathbf 1} + f_{ij}^k L^{cl}_k
+L^{cl}_k L^{cl}_l h_{ij}^{kl} +{\cal O}(1/N),
\label{algcc}
\end{align}
where $h_{ij}^{kl} $ is ${\cal O}(N^0)$ constant and
the Poisson bracket is
\begin{align}
-i \{ L^{cl}_i, L^{cl}_j \}_P=  \omega_{ij} {\mathbf 1} + f_{ij}^k L^{cl}_k
+L^{cl}_k L^{cl}_l h_{ij}^{kl}.
\label{Poic}
\end{align}

The commutator and the Poisson bracket should satisfy
the Jacobi identities.
For (\ref{Poic}), this implies
\begin{align}
\sum f_{ij}^l \omega_{kl}=0, \,\, 
\sum (f_{ij}^p f_{kp}^l +2 h_{ij}^{lp} \omega_{kp} )=0,
\label{ji}
\end{align}
where $\sum$ is taken over the cyclic permutations of  ${i,j,k}$.
Because only the unknown parameter are the anomalous dimensions
of the primary double trace operators constructed from the energy momentum tensors,
the last equation in (\ref{ji}) is expected to determine $h_{ij}^{kl}$.
Therefore, the classical limit of the CFT is uniquely 
determined by the Poisson bracket (\ref{Poic}) and 
the classical Hamiltonian $h_{cl}= \frac{1}{N^2} \hat{H}$
where $\hat{H}=D$.

We have seen that the classical limit of the $CFT_d$
may exist and is given by (\ref{Poic}).
The corresponding classical states for our case\footnote{
If we can neglect the multi trace operators, we can use the 
generalized coherent states based on Lie groups 
\cite{Kl} \cite{Pe}. 
It might be possible to incorporate the contributions of multi trace operators with minor modifications to this method,
we will explain the generalized coherent state based on (\ref{algc}) 
in the Appendix.
} 
can be taken as the generalized coherent state which is a 
deformation of the Harmonic oscillator coherent state.
The coherent state for the Harmonic oscillator is written as
$e^{-|\alpha|^2/2} \sum_{n=0}^\infty \frac{\alpha^n}{n!} \ket{n}$ 
where $\ket {n}$ is the normalized level $n$ state.
The deformed coherent state for the perturbed Harmonic oscillator 
is obtained from the coherent state by replacing $\ket {n}$
to the eigen state of the perturbed Hamiltonian which is deformed from $\ket {n}$.
This was used in \cite{BD, MV} although the commutator instead of the Hamiltonian is deformed for our case.
Thus the deformed coherent state labeled by $\{ \alpha_{A n lm} \}$ is
\begin{align}
\ket{ \alpha_{A n lm} }=
\prod_{A, n, l, m}  
\left[
e^{- |\alpha_{Anlm}|^2/2} \sum_{ {\cal N}_{A n lm} =0}^\infty \frac{(\alpha_{Anlm})^{ {\cal N}_{A n lm} } }{ {\cal N}_{A n lm} !}
\right]
\ket{ {\cal N}_{A n lm} },
\end{align}
where $\alpha_{A n lm}$ is ${\cal O}(N)$ constant.
This deformed coherent state may be regarded as the classical state. at least, for $\alpha_{Anlm}/N \ll 1$.

Until now, we have only used the properties of 
generic large $N$ gauge theories with the conformal symmetry
to derive the classical limit of the CFT.
It is expected that 
this classical system
is identified as the classical (Einstein) gravity on asymptotic $AdS_{d+1}$ space
because the classical system certainly reduces 
to the linearized gravity in the large $N$ limit taken in \cite{op}. 
We will show this identification is indeed correct in the next section.

Finally, we note that the this classical description is, of course, an approximation.
In particular, the Hilbert space we consider is the low energy approximation
and the classical approximation will be violated if the energy of the states are sufficiently large.
This bound of the energy for the classical approximation is (less than)  ${\cal O}(N^2)$ which is the degrees of freedom of the gauge theory.

\section{Classical gravity on asymptotic $AdS_{d+1}$ and the $CFT_d$}

In this section, we will consider the classical gravity on asymptotic $AdS_{d+1}$
in the Hamiltonian formalism, in particular using the Brown-York tensor.
We will see that the classical dynamics of  
generic large $N$ gauge theories with the conformal symmetry
is equivalent to the Einstein gravity on asymptotic $AdS_{d+1}$.
(We will explain the difference between the large $N$ expansion from the free theory 
and the classical limit in the Appendix \ref{free}.)

The Einstein-Hilbert action of the gravitational theory, with appropriate boundary terms, is
\begin{align} 
S_{grav}=\frac{1}{2 (l_p)^{d-1}} \int_{AdS_{d+1}} d^{d+1}x \sqrt{-\det g} 
\,
(R-2 \Lambda) +S_{GH}+S_{ct},
\end{align}
where $(l_p)^{d-1}=8 \pi G_N$, $\Lambda=-\frac{d(d-1)}{2 l_{AdS}^2}$
and we set the AdS scale $l_{AdS}=1$ in this section.
The Gibbons-Hawking term $S_{GH}$ is needed to allow the Dirichlet boundary condition
as a consistent boundary condition
and $S_{ct}$ is needed for making the action finite \cite{BaKr, deHaro}, although this term does not
play any role in the equations of motion of the classical dynamics which we concentrate
on this paper.
The metric of the vacuum solution is the $AdS_{d+1}$ metric:
\begin{align}
d s^2 = g_{\mu \nu}^{AdS} dx^\mu dx^\nu=-(1+r^2) dt^2 +\frac{1}{1+r^2} d r^2+ r^2 d \Omega_{d-1}^2,
\end{align}
where $0 \leq  r < \infty$, $-\infty < t < \infty$
and $d \Omega_{d-1}^2$ is the metric for the $d-1$-dimensional 
round unit sphere $S^{d-1}$.
Let us parametrize the metric as $g_{\mu \nu}=g_{\mu \nu}^{AdS}+h_{\mu \nu}$
and consider $h_{\mu \nu}$ as the varying fields.

First, we consider the free limit of the gravity, i.e. the
linearized gravity.
The e.o.m. of this limit was explicitly solved in \cite{IW} 
using the gauge invariant combinations. 
Because this is free theory,
we can easily see that 
in the (classical) Hamiltonian formalism, 
the results can be expressed as
\begin{align}
\{ a_{A n lm} , a^\dagger_{A' n' l' m'} \}_P=\delta_{A,A'} \delta_{n,n'} \delta_{l,l'} \delta_{m,m'},
\label{fP}
\end{align}
and 
\begin{align}
h_{cl}=\sum_{A, n, l, m, } \omega \, a^\dagger_{A n l m} a_{A n lm} ,   \,\,\, 
\label{fH}
\end{align}
where $(A, n, l, m) $ are the labels for the 
$L^+_{A \omega lm}$ where 
$\omega=d+l+r[A]-2+2n$ \cite{op}.
Here, we require that the only the normalizable modes of $h_{\mu \nu}$
are dynamical and 
the non-normalizable modes of $h_{\mu \nu}$ are set to be zero, which corresponds to fixing the boundary conditions.
This classical system is equivalent to the system with 
the Poisson bracket (\ref{Poi}) and the Hamiltonian 
as the dilatation if we neglect the terms proportional to $f_{ij}^k$, which are $1/N$ corrections to the free limit.

If we consider the full Einstein gravity,
we need to include the non-linear interactions of
the modes of $h_{\mu \nu}$.
Then, in the ADM formalism with an appropriate gauge fixing, we will have 
a Poisson bracket and a Hamiltonian 
which are modified by some $G_N$ corrections from
the ones for free cases, i.e.
(\ref{fP}) and (\ref{fH}).
It is possible to determine the corrections for the Hamiltonian explicitly, in principle, by the perturbation in $G_N$, but difficult practically.
However, we note that
the independent variables (which need not to be canonical)
may be labeled by same indices  $(A, n, l, m) $
for the free case and we will denote them as $a_{A n lm}$
even for the interacting case
because the interactions do not change the independent variables of the classical theory.\footnote{
Which values the indices  $(A, n, l, m) $ take are also same for the free case.
}

It is important to note that 
this system have the symmetry
corresponding to the isometries of $AdS_{d+1}$ with 
the metric $g_{\mu \nu}^{AdS}$, 
whose generators can be 
constructed by the Noether method or the asymptotic symmetry group, see, for example, \cite{reviewMa}.
Of course this symmetry is isomorphic to 
the conformal group of $CFT_{d}$
and includes the Hamiltonian as the dilatation.

Moreover, we can construct the 
analogue of the energy momentum tensor of 
$CFT_{d}$ in this system, $T_{\mu \nu}^{bndy} (x)$,
which is called
the boundary stress tensor, or the Brown-York tensor 
\cite{BrYo, BaKr, deHaro}.
The boundary stress tensor is defined as
\begin{align}
T^{\mu \nu}_{ bndy} (x) = \frac{2}{\sqrt{-\det g_{\mu \nu}^{(0)}}}
\frac{\delta S_{grav}}{\delta  g_{\mu \nu}^{(0)} (x) },
\end{align}
which is defined on 
the boundary of the asymptotically $AdS_{d+1}$, 
where $g_{\mu \nu}^{(0)}(x)$ is the boundary metric
and $\mu,\nu$ runs for the tangent directions of the boundary.\footnote{
To define the boundary stress tensor more precisely, we first introduce the 
IR cut-off for the radial direction $r$ and then remove the 
cut-off. 
The boundary metric is defined by removing the warped factor.
Details of the construction, see \cite{BaKr, deHaro, reviewMa}.
} 
It was shown that 
this boundary stress tensor is conserved, i.e. $\nabla_{\mu} T^{\mu \nu}_{ bndy} (x)=0 $ where $\nabla_{\mu} $ is the covariant derivative on the boundary
because of the diffeomorphism invariance of the action
\cite{BrYo, reviewMa}.
It was also shown that the trace of the boundary stress tensor vanishes for odd $d$ and a constant for even $d$ \cite{HeSk, BaKr, deHaro}.
This constant corresponds to the conformal anomaly 
if we assume the AdS/CFT.
Here, it is important that this constant is fixed if we fix the 
boundary metric, thus we can neglect this constant in the 
Poisson bracket.
In summary, the (traceless part of the) boundary stress tensor is 
symmetric, traceless and conserved.
Furthermore, 
this tensor will transform as a primary field by the
conformal symmetry transformation.
This is because 
the symmetry is associate with the diffeomorphism which 
induces the conformal transformation on the boundary metric 
$g_{\mu \nu}^{(0)}$.
With the Killing vector field for these diffeomorphisms,
the conformal symmetry generators are given by the 
boundary stress tensor as usual.
Thus, the boundary stress tensor $T^{\mu \nu}_{ bndy} (x)$ is regarded 
as an energy-momentum tensor of a $CFT_{d}$
where the commutators are defined by the Poissson bracket .

In the Hamiltonian or ADM formalism with a gauge fixing, which corresponds to a coordinate choice, 
the action $ S_{grav}$ and the boundary stress tensor 
$T^{\mu \nu}_{ bndy} $ are functions of the variables
$a_{A n lm}$ and $a^\dagger_{A n lm}$
where we fix the boundary metric $g_{\mu \nu}^{(0)} $
as the cylinder.
We can expand $T^{\mu \nu}_{ bndy} $ 
by the symmetric tensor harmonics 
and obtain the corresponding modes $L^{ bndy}_{A  lm} (|x|)$ 
and $\bar{L}^{ bndy}_{A n  lm} $ as in the previous section.\footnote{
This expansion might not be guaranteed to be valid in general.
However, at least, in the perturbation in $G_N$, we expect that such expansion is possible.
}
Thus, the modes $\bar{L}^{ bndy}_{A n  lm}$ 
are functions of  
$a_{A \omega lm}$ and $a^\dagger_{A \omega lm}$.
Conversely, the variables $\{ a_{A n lm} , a^\dagger_{A n lm} \}$
can be regarded as functions of $\bar{L}^{ bndy}_{A n  lm}$ 
where 
$l \ge 2$
and $n \in \z_{\ge 0}$ or $n+d+l+r[A]-2 \in \z_{\le 0}$,
i.e. the creation and annihilation operators,
(at least if the theory is close to the free theory)
because the number of the independent variables are same.
We can regard the map between these two as a field redefinition
although in order to obtain such a map explicitly we need to solve the equations of motion.
Note that the modes defined at the boundary can be equivalent 
to the whole bulk modes because the diffeomorphism gives the constraints
and the system in the $AdS$ space is like the system in a box \cite{Marolf, reviewMa}.

Therefore, the Poisson bracket algebra, which 
satisfies the Jacobi identities, of the boundary stress tensor
is same as the algebra of the energy momentum tensor
with the three parameters.
However, if we require the unitarity and the causality 
(and the sparseness of the spectrum which we already assumed),\footnote{
If we do not require these, then the Gauss-Bonnet gravity, 
and more generally the Lovelock gravity may correspond
to the theory with more parameters.
Note that the Lovelock gravity are the most general metric theory of gravity yielding second order equations of motion,
which is need to keep the number of the dynamical  variables.
For the Gauss-Bonnet gravity, the parameters are identified in \cite{Buchel}.
}
it have been shown \cite{CEMZ, Hartman} that there remains only one parameter $C_T$ 
in (\ref{opet})
which is ${\cal O}(N^2)$ for the gauge theory.
For the 
$T^{\mu \nu}_{ bndy} $, we can see $C_T \sim 1/G_N$.
This is because $T^{\mu \nu}_{ bndy} \sim 1/G_N$ by definition and
$\{ h_ , \dot{h} \}_P \sim G_N  $ where $h$ is the some component of $h_{\mu \nu}$,
schematically.
We can fix  the coefficient of this relation by computing above precisely
for the free theory limit and the result should agree with 
the result assuming the AdS/CFT correspondence as
$C_T =  \frac{d+1}{d-1}  \frac{\Gamma(d+1)}{\pi^{d/2}\Gamma(d-1)} \left(\frac{l_{AdS}}{l_p} \right)^{d-1}$ \cite{Buchel}.
The Hamiltonian of this gravitational system can be identified as the dilatation 
in the conformal symmetry. 
Thus, we conclude that 
the classical limit of the generic large $N$ gauge theory with conformal symmetry 
is 
the classical Einstein gravity on asymptotic $AdS_{d+1}$
because the Hamiltonian and the Poisson bracket are same.

%
%
%
%
%
%
%
%
%

\section*{Acknowledgments}

S.T. would like to thank  
Kanato Goto,
Shigeki Sugimoto and Sotaro Sugishita
for useful discussions.
S.T. would like to thank  
Yu Nakayama
for important comments and discussions.
This work was supported by JSPS KAKENHI Grant Number 17K05414.

\vspace{1cm}


\appendix

\section{Classical limit of $AdS/CFT$ and coherent state}

\label{free}

In this section, we will explain how to take the classical gravity limit
in terms of the large $N$ CFT assuming the AdS/CFT correspondence 
although this appendix is not used in the main parts of this paper.
We will also explain the (generalized) coherent states,
which are not used in the main parts of this paper.

Let us consider the following metric around the AdS space:
\begin{align}
ds^2=g_{\mu \nu} dx^\mu dx^\nu={(l_{AdS})}^2 
\left( {g}^{AdS}_{\mu \nu} dx^\mu dx^\nu 
+ h_{\mu \nu} dx^\mu dx^\nu \right),
\end{align}
where $l_{AdS}$ is the AdS (length) scale and $h_{\mu \nu}$ is the fluctuation around
the AdS space.

The action of the gravitational theory in the derivative expansion is, schematically, 
\begin{align} 
S_{grav} = \frac{1}{2 (l_p)^{d-1}} \int d^{d+1}x \sqrt{-\det g}
\left( -\frac{d(d+1)}{2 l_{AdS}^2}+R+\alpha_1 D^2 R + \alpha_2  l_{AdS}^2 R^2+
\cdots \right), 
\end{align}
where $(l_p)^{d-1}=8 \pi G_N$, $\alpha_i$ are dimensionless constants and 
$x^\nu$ and $h_{\mu \nu}$ are also dimensionless.
We assume that the AdS space is the solution of the e.o.m. of this action, 
as for the Gauss-Bonnet gravity or ${g}^{AdS}_{\mu \nu} $ is modified from the AdS metric
to be the solution with the higher derivative terms.

Let us define the dimensionless parameter $N$ as $N^2 =\left( l_{AdS} \over l_p \right)^{d-1}  $, then,
the action for the fluctuation is, schematically, given by
\begin{align} 
S=&  N^2 
\int d^{d+1}x \sqrt{-\det (g^{AdS})}  
\nonumber \\
& \,\, \times \left( -f_0(h) h^2+f_1(h) (Dh)^2+\alpha_1 f_2(h) (DDh)^2 + \alpha_2  f_3(h) (Dh)^4+
\cdots \right), 
\end{align}
where we abbreviated the various contractions of the indices 
and $f_i(h)$ is a function such that $f(h=0)$ is finite.

There are several choices for large $N$ limits.
One is the perturbation around the AdS geometry or 
the liner approximation.
For this, we will normalize $h_{\mu \nu}$ such that the 
kinetic term will be the canonical one.
Thus, we need to take $\tilde{h}_{\mu \nu} \sim N h_{\mu \nu}$ 
small, but finite.
If the higher derivative terms vanish, i.e. $\alpha_i \rightarrow 0$, in the large $N$ limit, 
we have the free theory of $\tilde{h}$ in the leading order in this large $N$ limit, 
with the spectrum given in 
\cite{IW} corresponding to the energy momentum tensor of the CFT \cite{op}.
Thus, $\tilde{h}$ is directly related to the creation/annihilation operators
of the free theory.
The sub-leading terms are interactions which include the terms in
$R$, $D^2 R$ and so on. In general, this $1/N$ expansion with $\tilde{h}$ include the classical and the quantum gravity effects.

Another choice of the large $N$ limit is the classical limit
where $h_{\mu \nu}$ is finite and $N^2$ is regarded as $1/\hbar$.
This means that the v.e.v. of the creation/annihilation operators,
i.e. $\tilde{h}$, should have ${\cal O}(N)$ values.
Note that this large $N$ limit contains the whole interactions 
of the Einstein gravity as a leading term, which are
non-leading term in the previous $1/N$ expansion with $\tilde{h}$.
In this paper, we will consider this large $N$ limit in the CFT.

\subsection{Coherent states for the linearized gravity}

We can consider classical states which is very close to the vacuum,
i.e. $h_{\mu \nu}= {\cal O}(N^0)$, but $ h_{\mu \nu} \sim \epsilon \ll 1$.
This is the linear approximation of the Einstein gravity (at least if $\alpha_i \rightarrow0 $
in the large $N$ limit). 
Here, we describe the coherent states for this approximation,
which have been considered in \cite{c0, c1,c2}.

Let us remember the coherent states for the free field (harmonic osscilator).
The coherent state for it is defined as 
$\ket{\alpha}=e^{\alpha a^\dagger -\bar{\alpha} a} \ket{0}$,
where $[a,a^\dagger]=1$ and $a \ket{0}=0$,
which satisfies $a \ket{\alpha}= \alpha \ket{\alpha}$
and $\int_{\mathbf C} d^2 \alpha \ket{\alpha} \bra{\alpha} =\pi$.
The overlap between the coherent state and 
the normalized energy eigen state $\ket{n}=\frac{1}{n!} (a^\dagger)^n \ket{0}$
is given by 
\begin{align}
\langle n \ket{\alpha} =e^{-\frac12 |\alpha|^2} {\alpha^n \over \sqrt{n !} }
=e^{-\frac12 |\alpha|^2+f(n)+{\cal O}(\ln n)},
\end{align}
where $f(n)=-\frac12 n(\ln n -2 \ln \alpha -1)$.
Thus, particle numbers where dominant contributions comes from 
for the coherent state is $n \sim \alpha^2$ 
because $\frac{\partial f(n)}{\partial n}=0$ at $n=\alpha^2$.

Therefore, the classical state  which is very close to the vacuum in the CFT 
is 
\begin{align}
\ket{\beta^i} = e^{N \epsilon \sum_i (\beta^i a^\dagger_i - \bar{\beta}^i a_i)} \ket{0},
\label{chf}
\end{align}
where $\beta^i ={\cal O}(N^0)$ are complex constants, 
$\epsilon$ is a small parameter
and $a^\dagger_i$ is the creation operator (\ref{ao}). 
This state is expected to be the coherent state 
$a_i \ket{\beta^i}  \sim N \epsilon \beta^i \ket{\beta^i}$,
where we assumed 
$[a_i,a^\dagger_j]=\delta_{ij} + {\cal O}(N^{-1})$
and the ${\cal O}(N^{-1})$ terms are neglected in the small $\epsilon$ limit.
The time evolution (in the Schr\"{o}dinger picture) is given by 
$i {\partial \over \partial t} \ket{\beta^i} = \hat{H} \ket{\beta^i} =D \ket{\beta^i}$
where $D$ is the dilatation operator and 
$[D,a_i]= \omega_i a_i$.
Thus, 
the solution is 
$\ket{\beta^i(t)} = e^{N \epsilon \sum_i (\beta^i(t) a^\dagger_i - \bar{\beta}^i(t) a_i)} \ket{0}$,
where ${\partial \beta_i(t) \over \partial t}=\omega_i \beta_i(t) $.\footnote{
We can replace $\ket{0}$ to any state in this. 
Indeed, $a_i$ is assumed to be diagonalized by the Hamiltonian, i.e.  $D$, thus
$a_i$ is the solution of the e.o.m.
}


\subsection{Coherent states for (\ref{algc})}

For the Lie algebra (\ref{algc}),
we can use the 
generalized coherent states based on Lie groups 
\cite{Kl} \cite{Pe}.
In this section, we assume (\ref{algc}) 
instead of (\ref{algcc}).
We basically follows \cite{Yaffe} to consider 
the generalized coherent states and their properties.\footnote{
There could be some differences between the large $N$ limit
taken in \cite{Yaffe}  and this paper.
The large $N$ limit in \cite{Yaffe} seems to 
correspond to the free limit because the factorization 
of the correlation functions were discussed.
Here, (\ref{algc}) include the $1/N$ corrections which violate 
the factorization properties.
}
First, we define the coherent group $G$
which consist of the following unitary operators:
\begin{align}
\hat{U}=e^{i (c^i L_i + c N^2 {\mathbf 1}  )}
=e^{i N^2 (c^i L^{cl}_i + c {\mathbf 1}  )},
\label{uni}
\end{align}
where $L_i$ are taken to be Hermite and $c^i, c$
are $N$-independent real constants.
We also define the isotropy subgroup $H$ of the coherent group whose element $\hat{V}$ satisfies 
$\hat{V} \ket{0} = \ket{0}$ up to a phase factor,
i.e. $\hat{V} =e^{i (c^i L_i^{iso} + c N^2 {\mathbf 1}  )}$.

Then, the generalized coherent states are defined
by $\hat{U} \ket{0}$ with parameters $c^i, c$ in (\ref{uni}).\footnote{
Because the Lie algebra is infinite dimensional,
we need to require some properties for $c_i$ such that 
the coherent state is well-defined, in particular, 
the state should have a finite energy.
A simple requirement for this is that only a finite number of 
$c_i$ do not vanish. 
This will be too strong condition and it is desirable to 
find an appropriate condition although we just assume 
the state is well-defined 
in this paper.
}
Using the Baker-Campbell-Hausdorff formula,
we can rewrite it as
\begin{align}
\hat{U} \ket{0} =C e^{\alpha^i L_i^{+} }\ket{0},
\end{align}
where $\alpha^i$ are some complex functions of $c^i, c$ which are 
$N$-independent and $C= e^{i N^2 \theta}  | e^{ \alpha^i L_i^{+} }\ket{0}|^{-\frac12}$ where 
$\theta$ is a $N$-independent real complex function of $c^i, c$ .
Thus, generalized coherent states are parametrized by $\alpha^i$ 
and the states 
with same $\alpha^i$ should be identified.\footnote{
The coherent states are parametrized by the the coadjoint orbit as we will see later.
}

We can also see that the classical operators defined in \cite{Yaffe} are the operators constructed from
$L^{cl}_i$.
In order to see this, let us consider two coherent states 
$\hat{U} \ket{0} =C e^{ \alpha^i L_i^{+} }\ket{0}$
and $\hat{U'} \ket{0} =C' e^{ ({\alpha'}^i L_i^{+} )}\ket{0}$
where $\hat{U},\hat{U}' \in G$.
Then, $\hat{U}^\dagger \hat{U'} \ket{0} = e^{i N^2 \theta_c} e^{ {\alpha}^i_c L_i^{+} }\ket{0}
\, |e^{ {\alpha}^i_c L_i^{+} }\ket{0}|^{-\frac12}$ is also a coherent state and
the overwrap of the two coherent states is given by
$\bra{0} \hat{U}^\dagger \hat{U'} \ket{0} 
=  e^{i N^2 \theta_c} \bra{0} e^{ {\alpha}^i_c L_i^{+} }\ket{0}
\, |e^{ {\alpha}^i_c L_i^{+} }\ket{0}|^{-\frac12}
= e^{i N^2 \theta_c} 
\, |e^{ {\alpha}^i_c L_i^{+} }\ket{0}|^{-\frac12}
$.
By the Baker-Campbell-Hausdorff formula, 
we expect the following rewriting:
$e^{ {\bar{\alpha}}^i_c L_i^{-} } e^{{\alpha}^i_c L_i^{+} }= 
e^{ {\beta}^i L_i^{+} } e^{  \gamma^i L_i^{iso} +  \phi N^2 {\mathbf 1} }
e^{{\bar{\beta}}^i L_i^{-} }$
where $\beta^i, \gamma^i, \phi $ are some $N$-independent constants
determined by $\alpha^i$.
Using this expression, we have ${\mathcal Re} (\ln \bra{0} \hat{U}^\dagger \hat{U'} \ket{0} )= -N^2 \phi/2$
where ${\mathcal Re} $ means the real part.
Then, if $\alpha_c^i $ is non zero for some $i$,
$\phi >0 $ because $|\bra{0} \hat{U}^\dagger \hat{U'} \ket{0}|<1 $.
Similarly, we can also show that 
${\mathcal Re} (\ln \bra{0} \hat{A}_{cl} \hat{U}^\dagger \hat{U'} \ket{0} )= -N^2 \phi/2 +{\cal O}(N^0)$
where $\hat{A}_{cl}$ is constructed from $L^{cl}_i$ without $N$-dependent coefficients.
This means that  ${\mathcal Re} (\ln \bra{0} \hat{U}^\dagger \hat{A}_{cl}  \hat{U'} \ket{0} )= -N^2 \phi/2 +{\cal O}(N^0)$
because
$\hat{U} \hat{A}_{cl} \hat{U}^\dagger = \hat{\tilde{A}}_{cl}$
where $\hat{\tilde{A}}_{cl}$ also is an operator constructed from $L^{cl}_i$ without $N$-dependent coefficients.

The classical operators defined in \cite{Yaffe} are the operators, say,  $\hat{A}$ such that
$\bra{0} U^\dagger \hat{A} U \ket{0} / \bra{0} \hat{U}^\dagger \hat{U'} \ket{0}$ is finite in the $N \rightarrow \infty$ limit.
Therefore, we find that the classical operators are indeed the operators constructed from $L^{cl}_i$.

We can also easily show that two coherent states, 
$\hat{U} \ket{0} =C e^{(\alpha^i L_i^{+} )}\ket{0}$
and $\hat{U'} \ket{0} =C' e^{(\alpha'^i L_i^{+} )}\ket{0}$,
are classically equivalent \cite{Yaffe}, which means 
$\bra{0} \hat{U}^\dagger \hat{A}_{cl} \hat{U} \ket{0}=\bra{0} \hat{U}'^\dagger \hat{A}_{cl} \hat{U}' \ket{0} $ for any $\hat{A}_{cl}$,  
if $\alpha^i=\alpha'^i$.
Even if $\alpha^i \neq \alpha'^i$, two states can be classically equivalent.
Including such identification, the coherent states are parametrized by the coadjoint orbit.
We will shortly explain this below.
First, denoting $g$ as the Lie algebra of the coherent group $G$, 
we can define the dual space $g^*$ whose elements are linear functionals acting on $g$.
Then, the expectation values of $L_i^{cl}$ for a coherent state,
$\bra{0} \hat{U}^\dagger L_i^{cl}  \hat{U} \ket{0} $, can be regarded as 
an element $\zeta^{\hat{U}}$ in $g^*$ if we regard $L_i^{cl}$ as basis of $g$.
In particular, we will denote $\zeta^{{\mathbf 1}}$ as for the element corresponding to $\ket{0}$
for which the components are given by $\zeta^{{\mathbf 1}}_i= \bra{0} L_i^{cl}   \ket{0}$. 
The coadjoint orbit $\Gamma$ is the set of $\zeta^{\hat{U}}$ in $g^*$ generating by $\hat{U}$.
Note that the expectation values of the operators constructed from $ L_i^{cl} $
are fixed by the $\zeta^{\hat{U}}$ because of the factorization of the expectation values \cite{Yaffe}.
Thus, classical equivalence class of the coherent states 
are identified as the coadjoint orbit\footnote{
The coadjoint orbit is parametrized by the possible (expectation) values of $L_i^{cl} $,
although $L_i^{cl} $ for $L^{iso}$ are not independent.
}
and 
then, the classical phase space is identified as the coadjoint orbit.

There are some requirements \cite{Yaffe} such that
the classical limit considered here is indeed behaves as the classical dynamics.
The one is the irreducibility of the representation 
of $G$ and this is satisfied because we assumed 
(\ref{state}) are all independent and our algebra reduced to 
the free harmonic oscillators, i.e. Heisenberg algebras,
for the small $c^i, c$.
It also required that 
if $\bra{0} U^\dagger \hat{A} U \ket{0}=0$
for any $U,U' \in G$, then $\hat{A} =0$.
This is also satisfied.
The requirement about the overwrap between 
the coherent states was already shown to be satisfied above.
The last requirement is about the Hamiltonian.
The classical Hamiltonian $h_{cl}$ is given by 
$h_{cl}= \frac{1}{N^2} \hat{H}$, where
$\hat{H}=D$ is the Hamiltonian in our theory, 
and $h_{cl}$ is indeed the classical operator.
Here, in the classical limit $h_{cl}$ is regarded as a function on the coadjoint orbit.
Thus, all the requirements are satisfied and we conclude that 
the (classical) equations of motion for a function $f$ on the coadjoint orbit,
which are parametrized by $L_i^{cl} $,  is
\begin{align}
\frac{d }{dt} f =\{ h_{cl}, f \}_P,
\end{align}
with the Poisson bracket (\ref{Poi})
for the classical limit of the CFT.

\end{document}